\documentclass[traditabstract]{aa}
\usepackage{txfonts,textcomp,natbib,amssymb,longtable}
\usepackage{graphicx}
\bibpunct{(}{)}{;}{a}{}{,} 
\newcommand{\Sauron}{\texttt{SAURON}}
\newcommand{\XSauron}{\texttt{XSauron}}

\begin{document}

\title{Explaining two circumnuclear star forming rings in NGC\,5248}

\author{T.\,P.\,R. van der Laan \inst{1,2}
\and E. Schinnerer \inst{1}
\and E. Emsellem \inst{3}
\and S. Meidt \inst{1}
\and G. Dumas \inst{2}
\and T. B\"oker \inst{4}
\and L. Hunt \inst{5}
\and S. Haan \inst{6}
\and C. Mundell \inst{7}
\and H. Wozniak \inst{8}
}

\institute{Max-Planck-Institut f\"ur Astronomie, K\"onigstuhl 17, 69117 Heidelberg, Germany; \textit{vanderlaan@iram.fr}
\and Institute de Radioastronomie Millimetrique (IRAM), 300 Rue de la Piscine, 38406 St. Martin d'Heres, Grenoble, France
\and European Southern Observatory, Karl-Schwarzschild-Str 2, 85748 Garching, Germany
\and European Space Agency, Keplerlaan 1, 2200AG Noordwijk, The Netherlands
\and INAF-Osservatorio Astrofisico di Arcetri, Largo E. Fermi 5, 50125 Firenze, Italy
\and CSIRO Astronomy and Space Science, ATNF, PO Box 76, Epping NSW 1710, Australia
\and Astrophysics Research Institute, Liverpool John Moores University, Birkenhead CH41 1LD, UK
\and Observatoire Astronomique de Strasbourg, CNRS UMR 7550, 67000 Strasbourg, France
}


\abstract{The distribution of gas in the central kiloparsec of a galaxy has a dynamically rapid evolution. Nonaxisymmetries in the gravitational potential of the galactic disk, such as a large scale stellar bar or spiral, can lead to significant radial motion of gaseous material from larger radii to the central region. The large influx of gas and the subsequent star formation keep the central region constantly changing. However, the ability of gas to reach the nucleus proper to fuel an AGN phase is not guaranteed. Gas inflow can be halted at a circumnuclear star forming ring several hundred parsec away. The nearby galaxy NGC\,5248 is especially interesting in this sense since it is said to host 2 circumnuclear star forming rings at 100\,pc and 370\,pc from its quiescent nucleus. Here we present new subarcsecond PdBI+30m CO(2-1) emission line observations of the central region. For the first time the molecular gas distribution at the smallest stellar ring is resolved into a gas ring, consistent with the presence of a quiescent nucleus. However, the molecular gas shows no ring structure at the larger ring. We combine analyses of the gaseous and stellar content in the central kiloparsec of this galaxy to understand the gas distribution and dynamics of this star forming central region. We discuss the probability of two scenarios leading to the current observations, given our full understanding of this system, and discuss whether there are really two circumnuclear star forming rings in this galaxy.}


\maketitle

\section{Introduction}
The short dynamical times in the central kiloparsec of galaxies lead to rapid changes in the gaseous and stellar distributions there. A large part of this secular evolution is driven by the inflow of new gaseous material from larger radii. The radial motion of gas in disk galaxies is dominated by gravitational torques on all but the smallest scales. These gravitational torques are generated by non-axisymmetric patterns in the disk, such as spiral arms, stellar bars, or ovals. However, continuous gas inflow from the larger scales down to the nucleus proper is not guaranteed under all conditions. For instance, gravitational torques from a large scale stellar pattern become much weaker with decreasing radius due to the increased mass influence of the (symmetric) stellar bulge.

In the case of a large scale stellar bar driving gas inward, the gas distribution is dominated by two spiral arms along the leading sides of the bar. In general, these spirals exist between the inner Lindblad resonance (ILR) and the corotation radius (CR) of the bar \citep[e.g.,][]{1978ApJ...221..521H,1979A&A....78..133B,Athanassoula1992b}. However, gas spirals that are generated by a stellar bar can exist beyond the ILR to the smallest radii to become nuclear spirals  \citep[e.g.,][]{2002ApJ...569..624P} under certain conditions. \citet{Englmaier2000} and \citet{Maciejewski2004II} show that the extent of nuclear spirals mostly depends on the gravitational potential, i.e. the relative strength of the large scale bar and stellar bulge, and the sound speed of the gas. When gas is kinematically warm (high sound speed), the pitch angle of the spiral will be large, resulting in a nuclear spiral which extents towards the nucleus. When gas is kinematically cold (low sound speed), the pitch angle is small, and the spiral will end in a circumnuclear ring. These rings act as reservoirs of gas from larger radii, and, as a result, form stars at a high rate.

When spiral arms end at a circumnuclear ring before reaching the nucleus proper, another mechanism might take over to funnel gas further inward. A nested stellar bar system, where one stellar bar exists within another, is one possible mechanism. This scenario was first proposed by \citet{1989Natur.338...45S}, and was followed up by many numerical studies \citep[e.g.][]{Maciejewski2002,2004ApJ...617L.115E,2010ApJ...719..622M}. One of the constraints on the stability of a nested stellar bar system is resonance overlap. Even then, gas flow into the region of influence of the inner bar is not always possible, but would be intermittent and dependent on the alignment of the two bars. 

\begin{figure*}
\includegraphics[width=17cm]{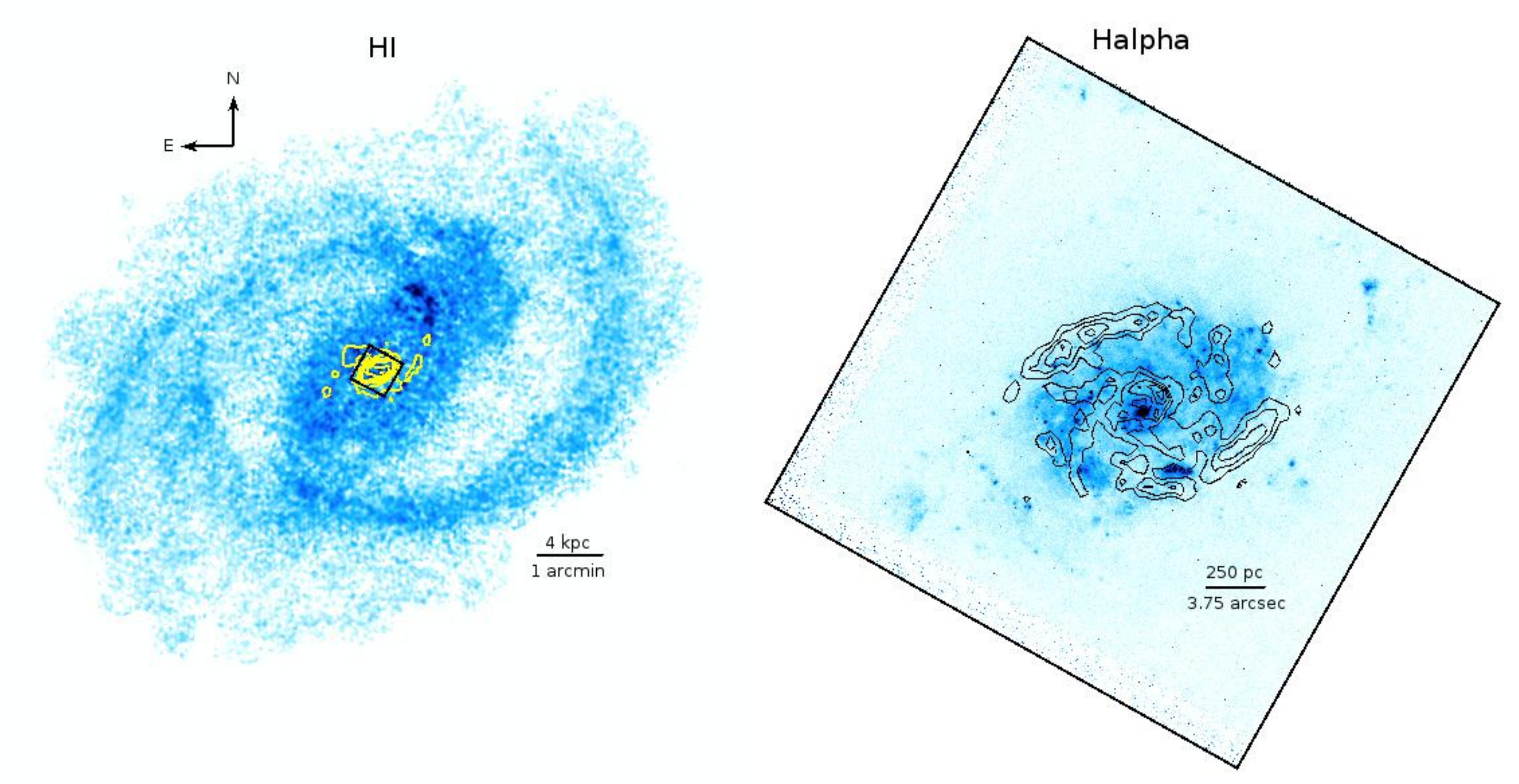}
\caption{Indication of the relevant scales in NGC\,5248. {\it Left:} the observed HI disk (color) goes out to 24\,kpc. The HI distribution appears dominated by two spiral arms and a large scale bar. BIMA SONG CO(1-0) observations are indicated with yellow contours in this panel. Molecular gas is only detected out to a 3\,kpc radius. The molecular gas is distributed in two gas arms and a central concentration. The FoV of the {\it HST} H$\alpha$ map is indicated with a black box. {\it Right:} a zoom-in of the {\it HST} H$\alpha$ map. The H$\alpha$ emission is dominated by the two circumnuclear star forming rings, at 100\,pc and 370\,pc from the nucleus. The CO(2-1) observations, presented in Sect. \ref{sect:CO21}, are overlaid in black contours.}
\label{fig:overview}
\end{figure*}

Nested bars are present in approximately 25\% to 40\% of the population of nearby barred galaxies \citep[e.g.][]{2002AJ....124...65E,2004A&A...415..941E}. These secondary bars are much smaller, and their size is no more than $\sim$12\% of the primary bar. Even in nearby galaxies, high angular resolution images are needed to detect them directly.

The NUGA survey \citep{NUGAstart} has further shown how multiple other patterns might work together to bring gas inward. Multiple configurations seem to be possible, which include lopsided disks (NGC\,4826: \citealt{2003A&A...407..485G}, NGC\,3718: \citealt{2005A&A...442..479K}, NGC\,5953: \citealt{Casasola2010}), bars and spirals (NGC\,4569: \citealt{Boone2007}, NGC\,2782: \citealt{Hunt2008}, NGC\,6574: \citealt{Lindt-Krieg2008}, NGC\,4579: \citealt{Garcia2009}, NGC\,6951: \citealt{Tessel}), and rings (NGC\,7217: \citealt{Combes2004}, NGC\,3147: \citealt{Casasola2008}, NGC\,1961: \citealt{Combes2009}).

Finally, for smaller rings, dynamical friction or viscose torques may lead to further gas flow towards the nucleus \citep{2008ASPC..396..325C}. As discussed in \citet{2005A&A...441.1011G}, the timescales for viscose torques only become comparable to those of gravitational torques at 100-200\,pc scales when strong density contracts (e.g. gas rings) are present. 

When no secondary pattern is present, or the ring is at too large radius for viscose torques to be effective, and a circumnuclear star forming ring forms, the ring is an effective barrier to further gas inflow \citep{Athanassoula2000,Maciejewski2004II,Haan2009,Tessel}. 

\begin{table}
\begin{minipage}{\columnwidth}
\centering
\caption{Global properties of NGC\,5248\label{tab:usefulvalues}}
\begin{tabular}{l l l}
\hline\hline
Parameter & Value & Reference \\
\hline
Type & SAB(rs)bc & (1)\\
\multicolumn{2}{l}{Dynamic Center}& \\ 
\,\,\,\,\,RA (J2000) & 13$^h$37$^m$32.0$^s$ & (2,3) \\
\,\,\,\,\,Dec (J2000) & 08\degr53\arcmin06.74\arcsec & (2,3)\\
Inclination Angle & 43.1\degr & (3,4)\\
Position Angle & 115\degr & (3,4)\\
Adopted Distance & 12.7 Mpc & (5)\\
Systemic velocity & 1153 km s$^{-1}$ & (3,4) \\
Mass H$_2$ (r$<$1.2\,kpc) & 4.0 $\times 10^8$ M$_{\sun}$ & (3) \\
Mass H$_2$ (r$<$300\,pc) & 1.0 $\times 10^8$ M$_{\sun}$ & (3) \\
\hline
\end{tabular}\\

\begin{flushleft}
\textbf{References:} (1) \citet{deVaucouleurs}, (2) NED, (3) this work, (4) \citet{Haan2008}, (5) \citet{2009AJ....138..323T}, (6) \citet{2002ApJ...570L..55J}
\end{flushleft}
\end{minipage}
\end{table}

In this paper, we study the case of the nearby barred galaxy NGC\,5248. This galaxy is said to host two circumnuclear star forming rings at 1.5\arcsec\, (100\,pc) and 6\arcsec\, (370\,pc) from its quiescent nucleus, in apparent contradiction with circumnuclear rings being effective gas barriers. Atomic and molecular gas has been detected at radii of less than 1.5\arcsec\, from the nucleus. NGC\,5248 is a nearby SAB(rs)bc galaxy at a distance of 12.7\,Mpc \citep{2009AJ....138..323T}, i.e. a spatial scale of 1\arcsec $=$ 61.6\,pc. Its global properties are listed in Table \ref{tab:usefulvalues}. 

Spiraling dust lanes between 6\arcsec\, (370\,pc) and 90\arcsec\, (5.5\,kpc) radius are seen in broad-band B and I-K images \citep[Figure 2 and 4a of ][]{Jogee2002}. The CO(1-0) distribution mapped by the BIMA SONG survey \citep{2003ApJS..145..259H} also shows some low-level CO emission at these positions (Fig. \ref{fig:overview}, left panel). \citet{2002ApJ...570L..55J} detected a large scale stellar bar in this galaxy, with a semi-major axis of 95\arcsec\, (5.9\,kpc), a PA of 135\degr\,, and a deprojected ellipticity of 0.44. Its pattern speed is approximated at 30\,km/s/kpc.

The two circumnuclear star forming rings are detected in H$\alpha$ emission \citep[Fig. \ref{fig:overview}, right panel,][]{2001MNRAS.324..891L,Maoz}. The outer one at 6\arcsec\, (370\,pc) is very broad, with a radius-to-width ratio of $\sim$2. The molecular gas distribution, as traced by CO(1-0) shows two spiral arms that are co-spatial with (part of) this circumnuclear ring. The orientation of these molecular spiral arms is similar to the larger dust spiral arms. In fact, \citet{Jogee2002} argue that both molecular gas, stars and dust spiral arms trace the same density wave, that thus winds up over 2$\pi$, from the outer edge of the large scale bar, past the larger circumnuclear ring, to the very center/smaller circumnuclear ring. The inner circumnuclear H$\alpha$ ring at 1.5\arcsec\, (100\,pc) is narrow and sharply defined. In the circumnuclear region all H$\alpha$ emission arises from these two rings, with a void between the two rings.

The primary goal of this work is to understand the make-up of the central 1.5\,kpc in this galaxy. We investigate the gas motions, and determine the stellar ages, to understand the gas distribution and dynamics, as well as the formation of the two circumnuclear star forming rings. To this end, new interferometric observations in the $^{12}$CO(2-1) line transition are presented in \S2. These observations have subarcsecond spatial resolution and resolve the inner circumnuclear ring in molecular gas for the first time. In \S3, gas morphology and non-circular motions in the gas are determined using archival HI and CO(1-0) data, together with our new CO(2-1) observations. In \S4 a thorough investigation of the ages of the stars in both rings is made. Work by \citet{Maoz} has identified and age dated star clusters observed from {\it HST} optical images before, but we repeat the age dating of the star clusters with updated stellar evolution models and our own $\chi^2$ fitting script. A \Sauron\, IFU data set \citep{2007MNRAS.379.1249D} is added to analyze the overall contribution of young stars in the central kiloparsec. In \S5 we discuss the star formation (history) and the gas dynamics in and between these two rings and discuss two possible scenarios that connect the different components. We summarize in \S6.

\section{Observations and data reduction}
\subsection{IRAM PdBI + 30M CO(2-1) data}\label{sect:CO21}
\begin{figure}
\resizebox{\hsize}{!}{\includegraphics[angle=-90]{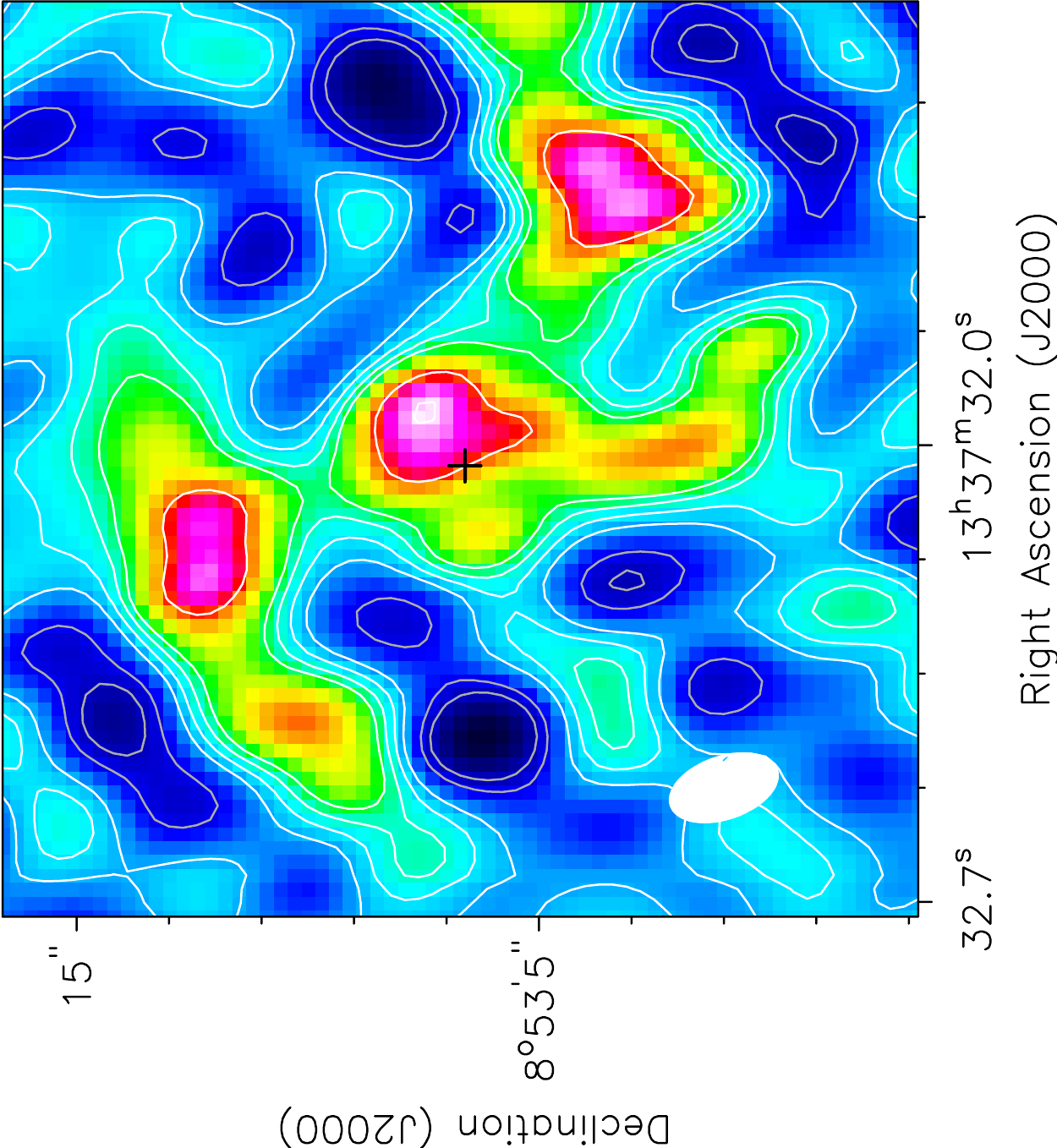}}
\caption{1\,mm continuum as observed with the PdBI Widex instrument. The telescope was positioned in D configuration. Contours give the -3$\sigma$, -2$\sigma$, 0, 2$\sigma$, 3$\sigma$ and then multiples of 5$\sigma$ flux levels (1$\sigma$= 1.8 mJy/beam). The galaxy center is indicated by a black cross. The beam size is 2.4\arcsec\,$\times$\,1.3\arcsec\, and is indicated in the bottom-left corner. Pixel scale is 0.30\arcsec. {\it A color version of this figure is available in the electronic version.}}
\label{fig:1mmcont}
\end{figure}

IRAM PdBI observations in ABCD configuration were carried out between January 2009 and December 2010 using the full 6-antenna array. The correlator was centered at 229.6547GHz, corresponding to a heliocentric velocity of 1153\,km\,s$^{-1}$, to observe the redshifted $^{12}$CO(2-1) emission line. Quasar 3C273 and MWC349 were used for flux calibration. During the observations in D-array, the WIDEX correlator was working in parallel with the narrow-band correlator, and simultaneously recorded the full 3.6\,GHz bandwidth. All observations were reduced using standard routines in the GILDAS/CLIC software package\footnote{http://www.iram.fr/IRAMFR/GILDAS}. A free addition to the CO observations with the WIDEX correlator at the PdBI is a 1\,mm continuum map. The WIDEX data were integrated over their full spectral range, excluding the CO(2-1) emission line region. The so obtained 1\,mm continuum map (Fig. \ref{fig:1mmcont}) has 128x128 pixels, a pixel scale of 0.3\arcsec/pixel, a beam size of 2.4\arcsec\, by 1.3\arcsec, and a PA of 18.8\degr. The highest emission peak reaches a 13$\sigma$ (1$\sigma$ = 1.8 mJy beam$^{-1}$) intensity. The 1\,mm distribution will be discussed in Sect/\,\ref{sect:agegradient}.

Single-dish observations for the central 44\arcsec\, (2.7\,kpc) by 44\arcsec\, (2.7\,kpc) were obtained with the IRAM 30m telescope on July 13 and 14, 2011. The 1\,mm receivers where tuned to 229.6553GHz. A bandwidth of 640\,MHz was covered by 512 channels with a width of 1.25\,MHz (2.6 km s$^{-1}$) at 1\,mm. The spacing between individual grid points was 4\arcsec\,. Typical system temperatures during the observations were 360\,K for the 1mm receivers. The data reduction was done using the GILDAS/CLASS software package.

The single-dish observations were used to compute the short spacing correction (SSC) and recover the large-scale low-level flux for the CO(2-1) line. The 30m observations were reprojected to the field center and frequency of the PdBI observations. The bandwidth coverage of the 30m data was resampled to match the velocity axis of the interferometric observations. A combined data cube was produced using the task `UV-short' in GILDAS.

A cleaned SSC image cube was produced using uniform weighting with the GILDAS/MAPPING software package. The beam size/resolution of the uniform weighted CO(2-1) data is given in Table \ref{tab:COcube}, as well as the r.m.s. noise per channel. The image cube has 1024 by 1024 pixels and 121 velocity channels, with a pixel scale of 0.1\arcsec/pixel and velocity bins of 5 km\,s$^{-1}$. CLEANing was done down to the 2$\sigma$ noise level with the assistance of CLEAN polygon regions defined individually for each channel with line emission present.

\begin{figure*}
\centering
\includegraphics[width=17cm, angle=-90]{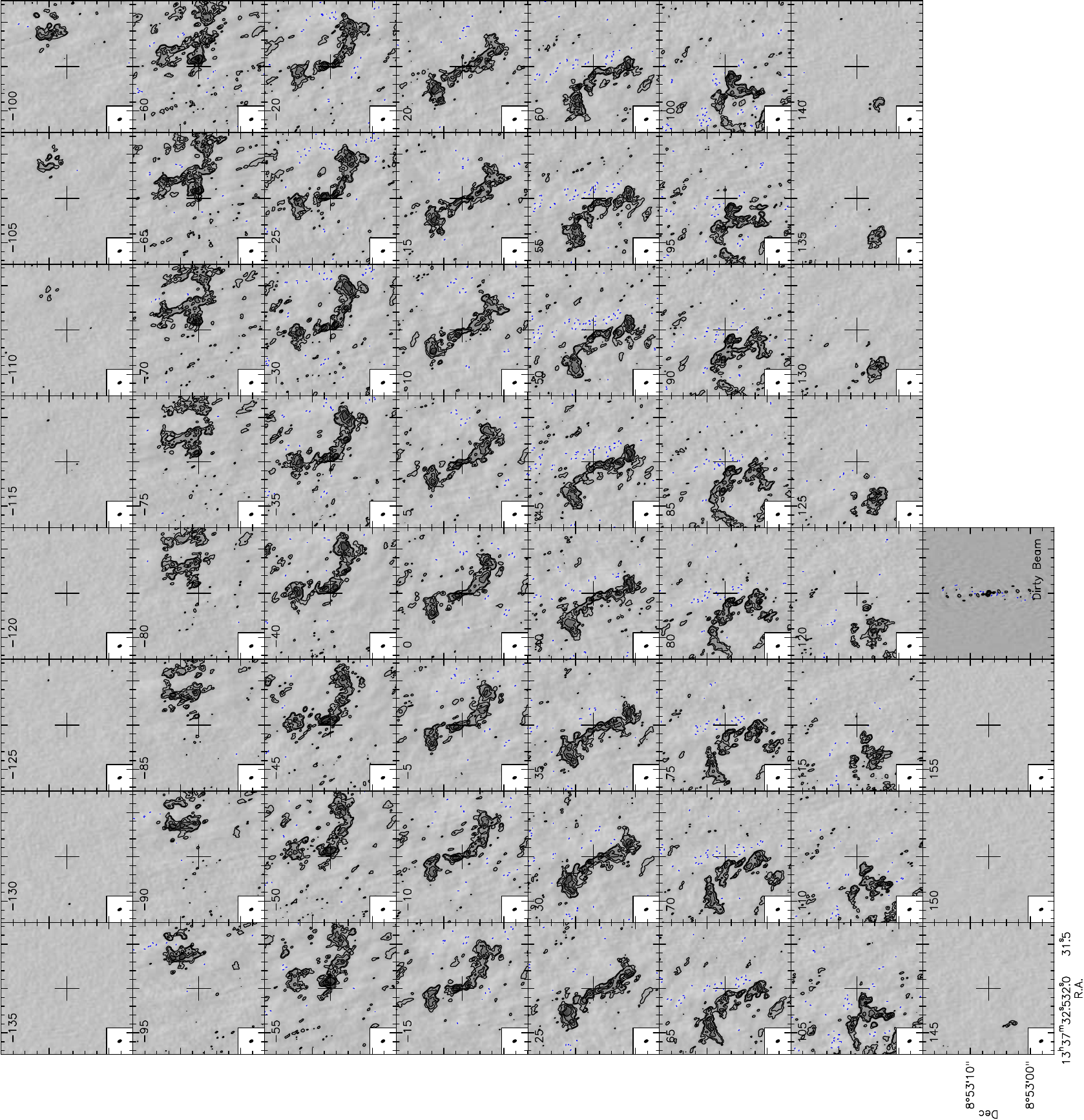}
\caption[Channel maps of the uniform weighted $^{12}$CO(2-1) SSC data cube]{Channel maps of the uniform weighted $^{12}$CO(2-1) SSC image cube. The size of the displayed channel maps is 22\arcsec\, by 22\arcsec\,, which is the size of the primary beam of the PdBI observations. The contours are at -3$\sigma$, 3$\sigma$, 5$\sigma$, 10$\sigma$, 15$\sigma$ and 25$\sigma$, with 1$\sigma$ = 2.8 mJy beam$^{-1}$. The velocity relative to the systemic velocity of the galaxy (v$_{sys}$ =1153 km s$^{-1}$) is indicated in the upper left corner. The phase center of the observations is indicated by a cross in each channel map. The clean beam (0.62\arcsec\, by 0.34\arcsec\,) is shown in the lower left corner of each channel map and the dirty beam is shown in the lower right panel.}
\label{fig:co_channel_un}
\end{figure*}

The channel maps of the cleaned image cube are presented in Fig. \ref{fig:co_channel_un}. CO(2-1) emission is seen in the channels [-130:145] km\,s$^{-1}$ relative to the 1153 km s$^{-1}$ systemic velocity. In the individual channels the emission is extended from north-east to south-west, in accordance with the position angle of 115\degr\, for this galaxy (see Table \ref{tab:usefulvalues}). The dirty beam is extended north-south due to the low declination of the source, which prevents a more circular population of the $(u,v)$-plane. The final clean beam has the size 0.62\arcsec\,$\times$\,0.34\arcsec\,, with a PA of 27.3\degr\,.

\begin{table}
\begin{minipage}{\columnwidth}
\caption{Key numbers of the CO(2-1) and 1\,mm continuum data}\label{tab:COcube}
\centering
\begin{tabular}{l c c c}
\hline\hline
 & Beam size & PA & r.m.s. \\
 & (\arcsec $\times$ \arcsec) & (\degr) & (mJy/beam)\\
\hline
CO(2-1) & 0.62$\times$0.34 & 27.3 & 2.8 \\
1mm cont. & 2.4$\times$1.3 & 18.8 & 1.8\\
\hline
\end{tabular}\\

\begin{flushleft}
{\bf Notes:} The CO(2-1) cube is constructed from PdBI ABCD + 30m observations, the 1\,mm cube only from the PdBI D configuration observations. Both cubes are imaged with uniform weighting.
\end{flushleft}
\end{minipage}
\end{table}

\subsection{Ancillary archival data}
For this study we also make use of the following data sets that have been presented in the literature before.

\begin{figure*}
\centering
\includegraphics[width=17cm]{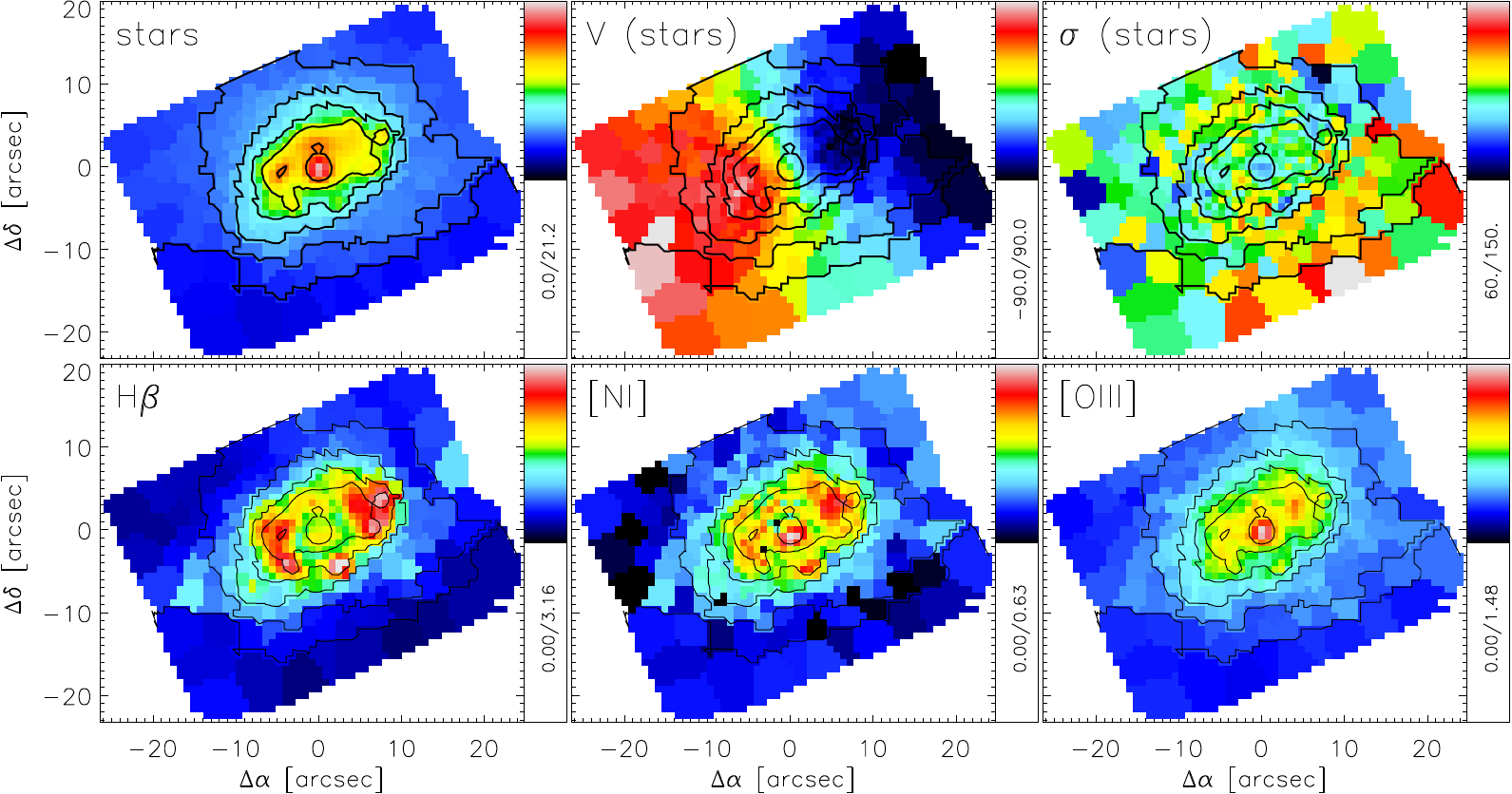}
\caption[NGC\,5248 Sauron maps]{{\it Top:} \Sauron\, maps of the stellar continuum, velocity field and velocity dispersion. {\it Bottom:} intensity distributions of the emission lines in the \Sauron\, spectral range; H$\beta$, [NI] and [OIII]. The maps are oriented with north pointing up, and east to the left.  The contours indicate the stellar flux in all panels and are included to guide the eye. {\it A color version of this figure is available in the electronic version.}}
\label{fig:sauron}
\end{figure*}

\subsubsection{\Sauron\ IFU data}\label{sect:SAURON}
NGC\,5248 was observed with the \Sauron\, instrument \citep{Bacon+01} at the William Herschel Telescope in March 2004 and maps were first presented by \citet{2007MNRAS.379.1249D}. The FoV of the \Sauron\ integral field unit (IFU) instrument is 33\arcsec\, (2.0\,kpc) $\times$ 41\arcsec\, (2.5\,kpc) with square 0.94\arcsec\, lenses. The final data cube is built out of 5 individual 30\,min exposures and has 0.8\arcsec$\times$0.8\arcsec\, spaxels. The spectral range covered is 4825 - 5280~\AA\,, with a 4.2~\AA\ resolution and spectral sampling of 1.15~\AA\,. Data reduction and flux calibration were done with the dedicated \XSauron\ software and is explained in detail in \citet{2007MNRAS.379.1249D}. Vonoroi binning was applied to increase the S/N to 40. For the central $\sim$9\arcsec\, (0.5\,kpc) radius the resulting Vonoroi bins are equal to the original spaxels (the S/N was already above 40).

The \Sauron\, spectral range (4825 - 5280)~\AA\, covers the stellar absorption lines H$\beta$, Fe5015, Mg$\beta$, and Fe5270, as well as the gaseous emission line H$\beta$, and the doublets [OIII] ($\lambda$ (4959,5007) $\AA$), and [NI] ($\lambda$ (5200,5202) $\AA$). Each spaxel is fitted with a combination of single-stellar population spectra (SSP) from the MILES library\footnote{http://miles.iac.es} \citep{Sanchez+06,2011A&A...532A..95F} and gaussian emission line profiles using the IDL-based routines pPXF \citep{CappEms04, Emsellem+04} and GANDALF \citep{Sarzi+06}. The kinematics of the emission lines were fixed to the [OIII] doublet. The combined use of pPXF and GANDALF takes into account the fact that H$\beta$ is present both in absorption and emission. 

Several of the resulting maps are shown in Fig. \ref{fig:sauron}. The top row in this figure shows the stellar continuum, stellar velocity field and stellar velocity dispersion. The distributions of the emission lines H$\beta$, [NI], and [OIII] are shown on the bottom row. To guide the eye, stellar flux contours are overlaid on all panels. The nebular emission lines are strongest in the outer circumnuclear ring, which is most clearly traced in H$\beta$ emission, but also in [NI] and [OIII]. Some H$\beta$ emission is also seen from the inner 2\arcsec\,, and seems to be not located at the nucleus, but at the inner circumnuclear ring. The [NI] line emission peaks at this same position at the inner ring. The ratio [NI]/H$\beta$ is an indicator of the intensity of the continuum fluorescent excitation connected to HII regions \citep{1999ApJ...527..474B,2010MNRAS.402.2187S,2012ApJ...757...79F}. Its value ($\sim$0.1) in the southern half of the inner ring implies recent, intense star formation. In the outer circumnuclear ring the positions of highest H$\beta$ and [NI] line emission intensity also are co-spatial, but the [NI]/H$\beta$ ratio is much lower ($\sim$0.03). The lower value would indicate that recent star formation is less intense in the outer ring. At the outer circumnuclear ring the emission of all three lines is strongest at the north-west and south-east quadrants of the ring. This could hint at an azimuthal star formation distribution. The stellar continuum does not appear particularly correlated with the emission line intensities, which suggests that the distribution of stellar light is not overall dominated by the youngest stellar population. Stellar continuum emission is higher at the northern section of the ring, than at the southern part. 

\subsubsection{HI}
NGC\,5248 was observed with the VLA in C and D array as part of a larger campaign \citep{Haan2008,Haan2009} in 2003 and 2004. We refer to that work for details of the observations and data reduction. The robust weighted data cube used in this work has a spatial resolution of 20\arcsec, with a 2\arcsec\, pixel size and a FoV of 34\arcmin (125\,kpc) $\times$ 34\arcmin (125\,kpc). The total HI flux measured in these observations is 81.3 Jy\,km\,s$^{-1}$, which corresponds to an atomic gas mass of 4.32$\times10^{9}M_{\sun}$ \citep{Haan2008}.

The HI distribution (Fig. \ref{fig:overview}, left panel) is dominated by a central bar-like structure (most likely a response to the 95\arcsec\, (5.9\,kpc) large scale stellar bar) and spiral arms. The HI intensity is somewhat higher within the central region, but within that there is little to no variation, nor is there an intensity peak towards the center. The morphology of the atomic gas does not trace spiral structure within the central region, except for a partial spiral arm in the north-west that continues into a CO arm at smaller radii. 

\subsubsection{BIMA SONG CO(1-0)}
The BIMA SONG survey \citep{2003ApJS..145..259H} was designed to systematically observe the CO(1-0) distribution in 44 nearby galaxies. NGC\,5248 was observed in May 1999. The total FoV of these data are 194\arcsec\, (12\,kpc) $\times$ 194\arcsec\,, (12\,kpc) the data have a resolution of 6\arcsec\,, and are short-spacing corrected. The resulting data cube and moment maps are publicly available from NED\footnote{NASA/IPAC Extragalactic Database}. A contour plot of the intensity distribution is included in Fig. \ref{fig:overview} (left panel).

The BIMA SONG intensity distribution shows that the molecular gas density, as traced by the CO(1-0) flux intensity, increases towards the center, with an `S'-like shape, which is resolved into two spiral arms in the higher angular resolution CO(1-0) maps by \citet{Jogee2002}. The main CO(1-0) emission structure extends out to $\sim$22\arcsec\, (1.4\,kpc) from the nucleus. At larger radii, two spiral arms are partly traced by the CO(1-0) emission. No CO(1-0) line emission is detected outside a 3\,kpc radius.

\subsubsection{HST}
In \citet{Maoz} a thorough detection of star clusters in the circumnuclear rings of NGC\,5248 is presented. This work was based on HST observations in 5 filters; F220W, F336W, F547M, F814W and F160W. A table with the observed fluxes for each star cluster in each band was available for download with the paper. We refer to that work for details on the data reduction and cluster finding. In total fluxes were derived for 507 star cluster candidates (i.e. point sources) in one or more filter bands. The radial distribution of candidates in the outer circumnuclear ring is wide, $\Delta$r\,$\sim$2\arcsec\,, especially in the north-west region of the outer ring. The distribution at the inner ring is very narrow, $\Delta$r\,$\leq$0.2\arcsec\,.

\section{Molecular gas in the center of NGC\,5248}
The existence of two circumnuclear star formation rings in NGC\,5248 is unexpected in relation to the understanding that circumnuclear rings are effective gas barriers. Especially since this galaxy is believed to have only one large scale stellar bar. Fortunately, the morphology and kinematics of the gas in the circumnuclear region can be used to show the (formation) relation between the two rings, if any, and other structures in the circumnuclear region.

\subsection{CO(2-1) morphology}
\begin{figure}
\resizebox{\hsize}{!}{\includegraphics[angle=-90,width=\columnwidth]{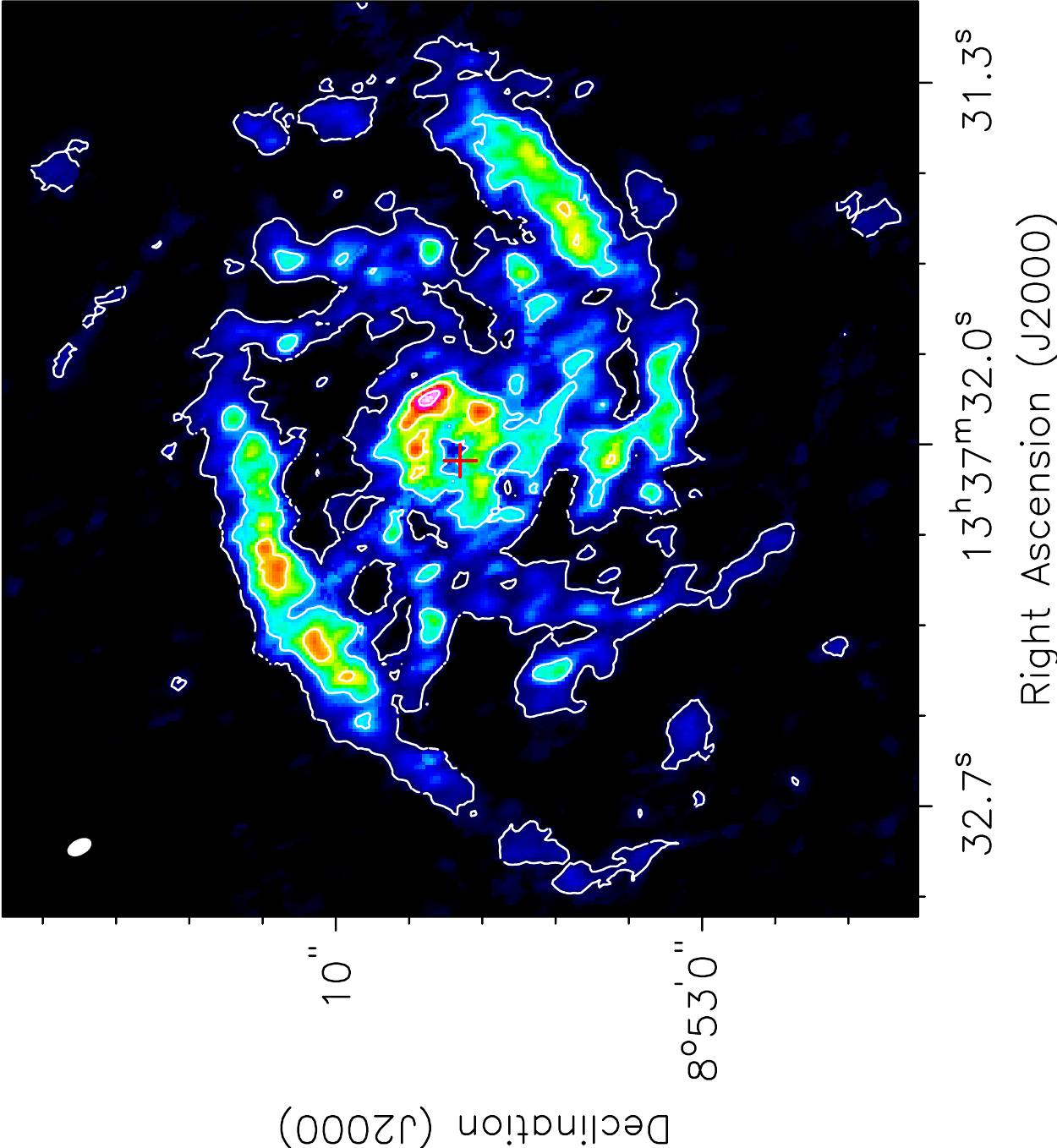}} 
\caption[$^{12}$CO(2-1) zeroth moment map of NGC\,5248]{Integrated CO(2-1) emission in uniform weighting for the inner 1.5\,kpc radius of NGC5248. The CO(2-1) emission has been integrated from -135 km s$^{-1}$ to 155 km s$^{-1}$ relative to the systemic velocity of 1153 km\,s$^{-1}$. Contours run from 5$\sigma$ in 10$\sigma$ steps (1$\sigma$ = 0.014 Jy beam$^{-1}$ km s$^{-1}$) The red cross indicates the position of the dynamic center. The beam size is shown in the upper left corner and corresponds to the values listed in Table \ref{tab:COcube}. {\it A color version of this figure is available in the electronic version.}}
\label{fig:co_0mom}
\end{figure}

The new CO(2-1) observations presented here have sub-arcsecond resolution and offer an unprecedented view of the molecular gas in the central region of NGC\,5248. The intensity distribution of the CO(2-1) line emitting gas is shown in Figure \ref{fig:co_0mom}. At 6\arcsec\, (370\,pc) from the nucleus two partial spiral arms are seen in the north and south. The location of the spiral arms is co-spatial with the location of the outer circumnuclear star forming ring. Inside that, the morphology shown in Fig. \ref{fig:co_0mom} is difficult to interpret. It can be explained as two continuing spirals that extend the two outer CO arms inwards for another 180\degr\,, or as two additional, fractured, lower-intensity arms originating east and west of the nucleus. At 1.5\arcsec\, (100\,pc) radius from the center we find a nearly-full ring. This location is consistent with the location of the inner circumnuclear star forming ring, and it is the first time that this inner ring is resolved in molecular gas. The CO(2-1) intensity reaches the highest values in this ring. Very little CO(2-1) emission is detected from the nucleus.

An integrated flux of 684 Jy\,km\,s$^{-1}$ is measured within the central 10\arcsec\, radius, of which 177 Jy\,km\,s$^{-1}$ ($\sim$26\%) are within the central 3\arcsec\, radius (the inner ring). This corresponds to a molecular gas mass of 4.0$\times10^{8}M_{\sun}$ and 1.0$\times10^{8}M_{\sun}$, respectively. These values are corrected for the helium abundance, and an assumed, constant, I$_{CO(2-1)}$/I$_{CO(1-0)}$ line ratio of 0.89 \citep{1993A&AS...97..887B} was used. Similar ratios are found for the central regions of other nearby galaxies, \citep{2012ApJ...761...41K,2012arXiv1212.1208S}. The Galactic X$_{CO}$ value of 2.2e20 cm$^{-2}$\,[K\,km\,s$^{-1}$]$^{-1}$ is also assumed\footnote{Recent work by \citet{2012arXiv1212.1208S} seems to indicate that this value is actually lower in the centers of galaxies. A ratio lower by a factor $x$, would lower the derived gas mass by a factor $x$.}. Table \ref{tab:usefulvalues} summarizes these and other relevant numbers for NGC\,5248. 

Comparison between the CO(2-1) emitting gas distribution presented here and the CO(1-0) distribution presented in \citet{Jogee2002} shows that CO(1-0) and CO(2-1) emission is both present in the two spiral arms that trace the outer circumnuclear star forming ring in the north and south. Inside the outer ring radius the two line emission distributions differ. The CO(1-0) emission flux is lower than in the spiral arms, while the CO(2-1) emission is brightest in this region. This could indicate that CO(2-1) is preferentially excited over CO(1-0) at these smaller radii \citep[this would imply a higher I$_{CO(2-1)}$/I$_{CO(1-0)}$ ratio for the inner ring, e.g.][and decrease the molecular gas mass in the inner ring with respect to the outer ring]{2012ApJ...752...87A}. The morphology of the CO(1-0) and CO(2-1) emission within the outer ring radius is also different. The CO(1-0) emission is distributed in an elongated structure parallel to the CO spiral arms that are co-spatial with the outer ring, with emission peaks at the inner circumnuclear ring (east and west) and extending low-level emission spurs at the outer circumnuclear ring raidus, whereas the CO(2-1) emission shows the low intensity spiral arms in the east and west at the outer ring radius and the distinct structure of the inner circumnuclear ring in the center. The different appearance in CO(1-0) and CO(2-1) line emission inside the large star forming ring could reflect changes in the excitation conditions from the outer to the inner ring, with young star formation, i.e. heating of the molecular gas, being a very likely explanation. Recall that in the \Sauron\, data (\S \ref{sect:SAURON}) a high [NI]/H$\beta$ ratio was seen at the inner ring, but not the outer ring.

\begin{figure*}
\centering
\begin{tabular}{c c}
\includegraphics[angle=-90,width=8.6cm]{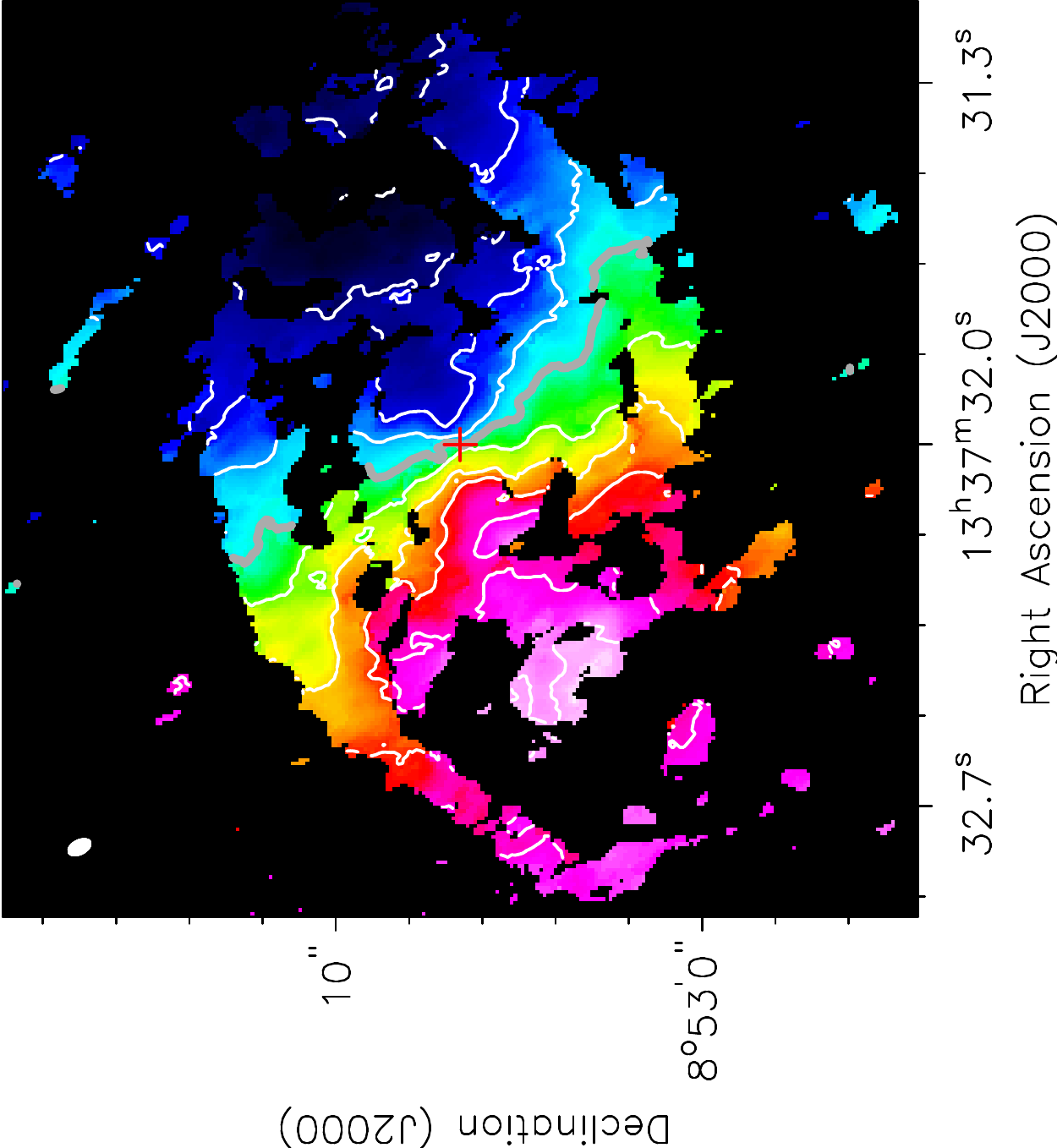}&
\includegraphics[angle=-90,width=8.6cm]{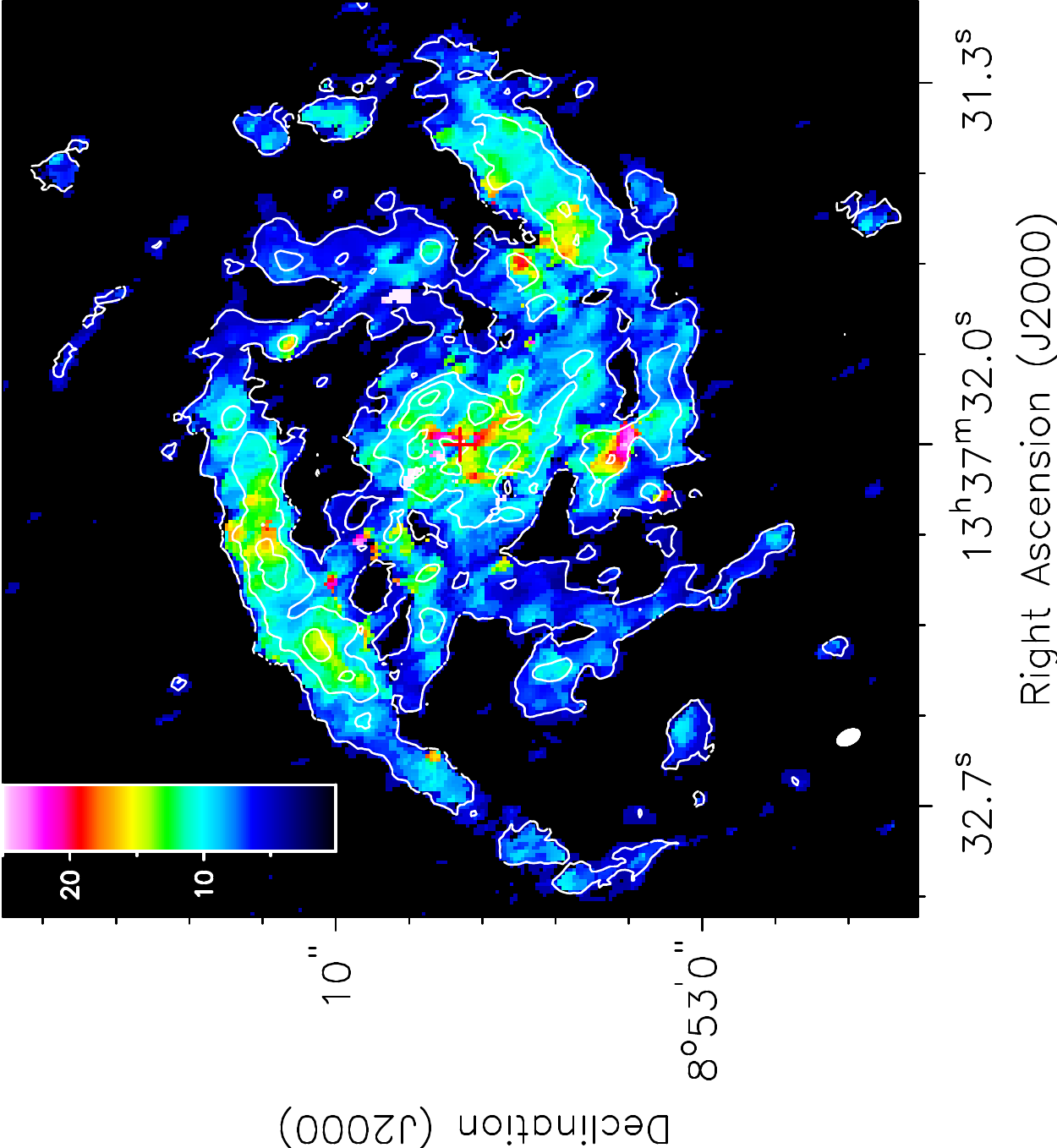}\\
\end{tabular}
\caption{{\it Left:} First moment map of the CO(2-1) emission. Contours: -150 to 150 km s$^{-1}$ with 25 km s$^{-1}$ steps, 0 km s$^{-1}$ thick gray contour. Velocities are relative to the systemic velocity, 1153\,km\,s$^{-1}$, with negative velocities in blue and positive velocities in red. {\it Right:} Second moment map of the CO(2-1) emission. The color bar shows the dispersion in km $^{-1}$. Contours are the integrated CO(2-1) emission in 10$\sigma$ steps starting at 5$\sigma$ (1$\sigma$ = 0.014 Jy\,beam$^{-1}$\,km\,s$^{-1}$). The dynamic center is given with a red cross in both panels. {\it A color version of this figure is available in the electronic version.}}
\label{fig:co_kin}
\end{figure*}

\subsection{CO(2-1) kinematics}
The first (velocity field) and second (velocity dispersion) moment maps of the CO(2-1) data are shown in Fig. \ref{fig:co_kin}. The velocity field of the CO(2-1) emitting gas is dominated by circular velocity; it shows the `spider diagram' contours of an inclined rotating disk. The dynamical center is equal to the photometric center, within the observational resolution. The kinematic axis of the velocity field of the CO(2-1) emitting gas and the stellar velocity field of the \Sauron\, observations are very similar. The CO(2-1) velocity range is broader than the \Sauron\, stellar velocity range, but the velocity dispersion is broader in the stars. This implies that the molecular gas is kinematically colder ($\sigma_{CO(2-1)}\sim$5-20\,km\,s$^{-1}$) than the stars, and shows more ordered motion. There is no offset between the stellar and gaseous disk.

Given that spiral arms are usually found at the leading side of a bar and that the blue-shifted velocities are in the west, NGC\,5248 must be rotating clockwise. This implies that the south-western part of NGC\,5248 is the near side of the galaxy, and the north-eastern part the far side. The orientation of the disk is important to be able to determine later on whether non-circular motions imply in- or out-flows.

The highest CO(2-1) velocity dispersions (15-20 km s$^{-1}$) are seen in the spiral arms at 6\arcsec\, (370\,pc) from the nucleus and inside the inner 4\arcsec\, (250\,pc), i.e. the inner ring region. Given the location, the higher velocity dispersions could be indicative of either enhanced turbulence due to recent star formation or non-circular motions caused by shocks or shear in the gas.

\begin{figure*}
\begin{tabular}{c c}
\includegraphics[angle=-90,width=8.6cm]{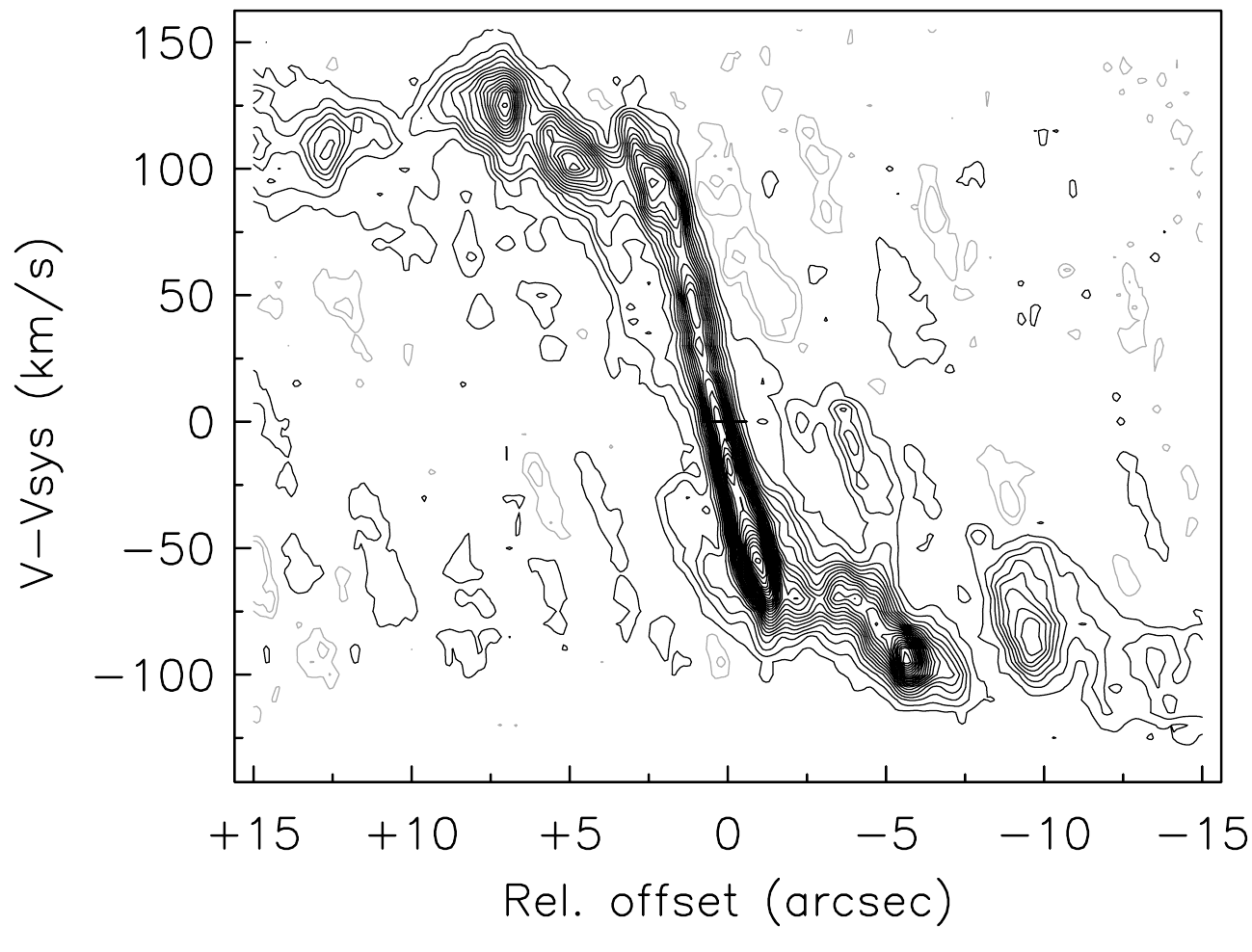} &
\includegraphics[angle=-90,width=8.6cm]{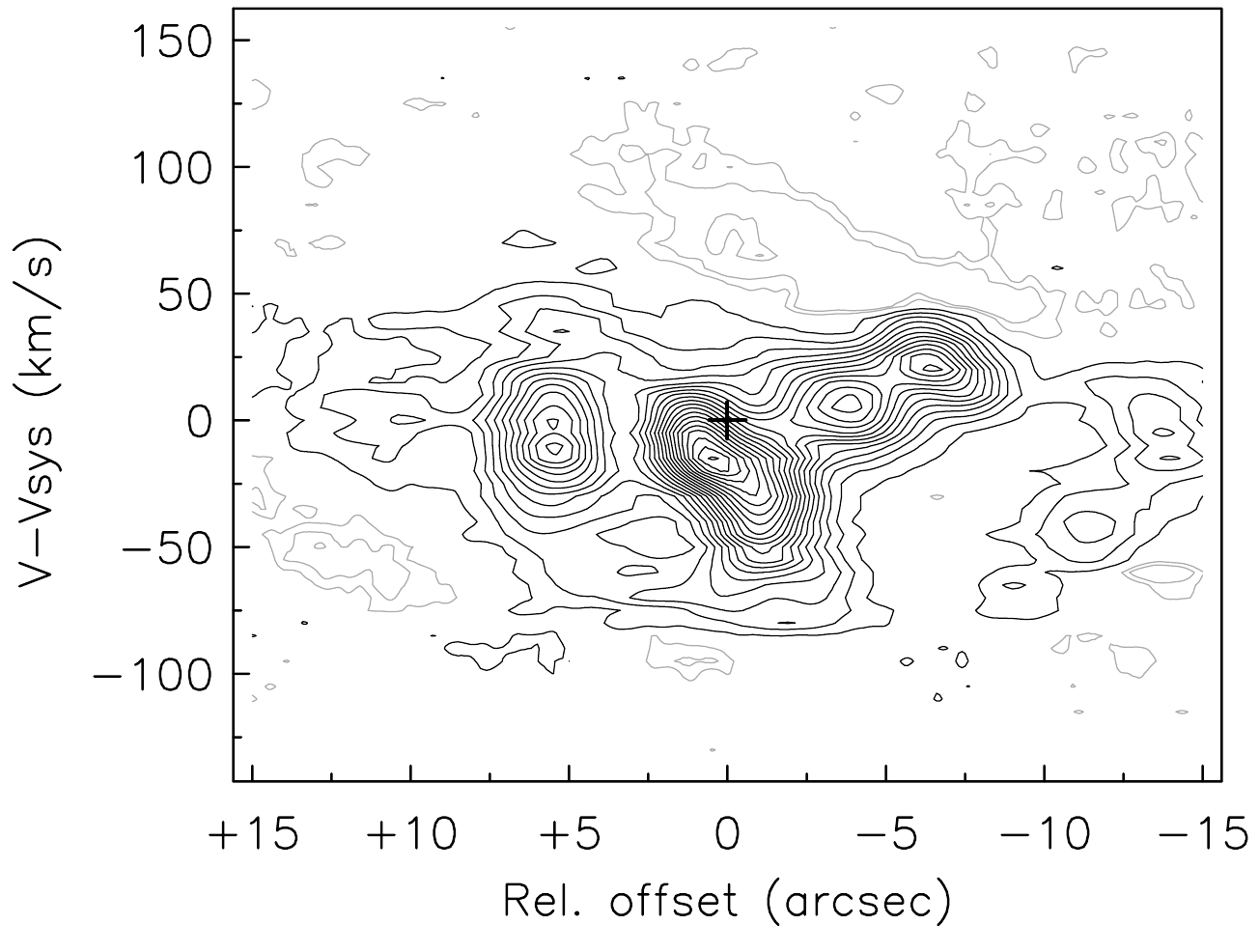} \\
\end{tabular}
\caption{{\it left:} Position-velocity diagram along the major axis (PA of 115\degr) through the dynamic center. The cut was made from the SE (positive offsets) to the NW (negative offsets). {\it right:} Position-velocity diagram along the minor axis (PA of 25\degr) through the dynamic center, from the NE (positive offsets) to the SW (negative offsets). In both panels, the contours are at 2$\sigma$, in 2$\sigma$ steps. The dynamic center is at (0,0).}
\label{fig:co_pvdiagram}
\end{figure*}

As a first investigation of the non-circular motion two position-velocity (pv) diagrams were extracted from the CO(2-1) data cube; one along the major kinematic axis (PA 115\degr\,) and one along the minor kinematic axis (PA 25\degr\,); see Fig. \ref{fig:co_pvdiagram}. The pv-diagram along the major axis shows evidence for solid body rotation for r$<$2\arcsec\, (130\,pc). The linear `pv' distribution along the major kinematic axis indicates that gas along this line is rotating with a single angular velocity. The turn-over of the velocity at larger radii ($\pm$ 4\arcsec) is not smooth, which implies that there might be some non-circular motions, or the presence of both x$_1$ and x$_2$ orbit families, belonging to a barred potential \citep[e.g.][]{2006MNRAS.370.1499A} present at this radius. More evidence for non-circular motions are found in the pv-diagram along the minor axis. If the gas was rotating with only circular velocities, then all velocities along the minor axis should be zero. This is not the case, in the r$<$2\arcsec\, (130\,pc) range there is some offset (up to -50 km s$^{-1}$). Also at larger radii deviations, up to 30\,km\,s$^{-1}$, along this axis are present, particular at -5\arcsec\,to -7\arcsec\, (that is, south of the nucleus). In both position-velocity diagrams, some low-level emission, away from the main emission structure, is visible. Due to the low declination of the source, and the orientation of the major axis, the beam side lobes were oriented in the same direction as the emission in individual channels. This made it very difficult to completely remove them. Here we see the remainder, which is only at the 2$\sigma$ level.

\subsubsection{Non-circular gas motions}\label{sect:noncirc}
We investigate the location of the non-circular motions in the gas in more detail. We use the GIPSY routine ROTCUR \citep{1992ASPC...25..131V,2001ASPC..238..358V} to construct a rotation curve spanning the entire galaxy out to a radius of 24\,kpc. We combine our CO(2-1) data with the BIMA SONG CO(1-0) and VLA HI data, so that this rotation curve has, at the same time, high resolution in the central kiloparsec.

A galaxies' observed line-of-sight velocity field can be decomposed into harmonic components, V$_{los}$ = V$_{sys}$ + V$_{circ}$ + V$_{rad}$ + higher order terms \citep{1993ApJ...414..487C,1997MNRAS.292..349S}, where V$_{sys}$ is the systemic velocity, V$_{circ}$ the pure rotational component to the velocity, and V$_{rad}$ the pure radial component to the velocity. The ROTCUR determines these velocity components, V$_{sys}$, V$_{circ}$, and V$_{rad}$, by fitting tilted rings (parameterized with inclination, PA, and dynamical center) as a function of radius to the observed line-of-sight velocity field. (Higher order terms are not present in ROTCUR.) The systemic velocity V$_{sys}$ (1153km/s) as well as the dynamical center and inclination (43.1\degr\,) of NGC\,5248's disk were determined, and then fixed, in an initial run of ROTCUR on the HI data alone, since the HI observations have the larger extent and are therefore less sensitive to the expected radial motions in the central kiloparsec. Further investigation then showed no significant change in the PAs of the tilted rings fitted to the different gas tracers as function of radius, the values were in accordance with the major axis orientation, thus the PA was fixed to 115\degr\, (the major axis orientation). Finally, we also kept V$_{rad}$ at zero, and let ROTCUR fit the only remaining variable, V$_{circ}$. 

\begin{figure*}
\centering
\begin{tabular}{c c}
\includegraphics[height=7cm]{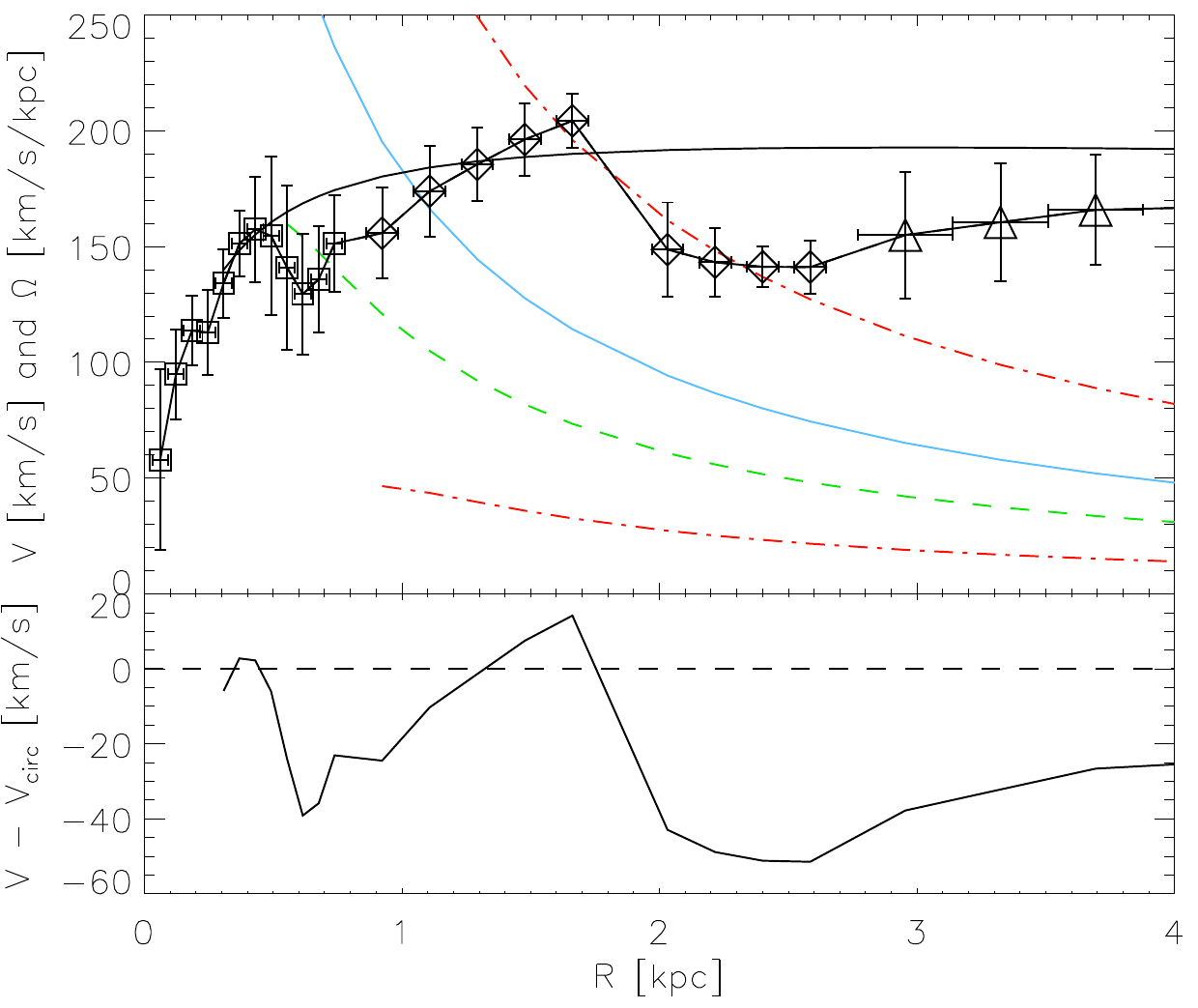}&
\includegraphics[height=7cm]{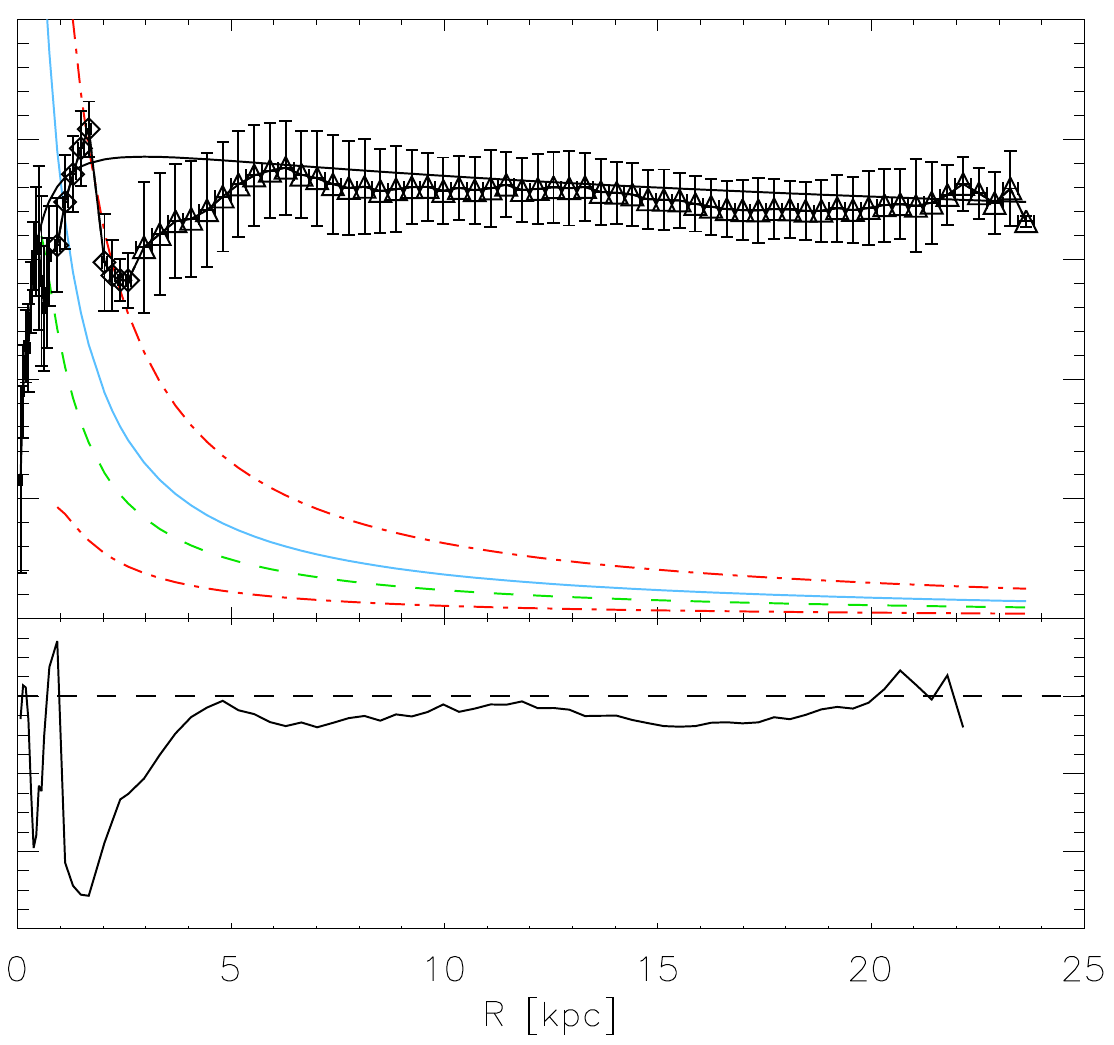}\\
\end{tabular}
\caption[Gas rotation curve of NGC\,5248]{Gas rotation curve of NGC\,5248, spanning the inner 4\,kpc ({\it left}) and the entire disk out to 24\,kpc ({\it right}). The rotation curve is based on our CO(2-1) (squares), as well as BIMA SONG CO(1-0) (diamonds) and VLA HI (triangles) data, giving both high spatial resolution and large radial coverage. A circular velocity curve is fit using equation \ref{eq:rot}, solid line. Frequency curves of $\Omega$ (solid, blue), $\Omega +\kappa/2$, $\Omega -\kappa/2$ (both dash-dot, red) and $\Omega -\kappa/4$ (dashed, green) are also presented. The bottom panels show the difference between circular velocity model curve and ROTCUR derived rotation curve. (See text for further details.)}
\label{fig:rotcurve}
\end{figure*}

From the center to the edge of the PdBI FoV, the CO(2-1) velocity was sampled around the major axis every 1\arcsec\, out to 11\arcsec\, (620\,pc), excluding a 20\degr\, wedge around the minor axis. Similarly, the BIMA SONG velocity was sampled every 3\arcsec\, out to 42\arcsec\, (2.6\,kpc). Finally, the HI velocity field was sampled every 6\arcsec\, out to 6.5\arcmin\, (24\,kpc). The derived velocities match very well at the radii where the different datasets overlap. Fig. \ref{fig:rotcurve} shows both a zoom of the inner 4\,kpc ({\it left}) and the full rotation curve ({\it right}).

The resulting velocity curve now under-predicts the `true' circular velocity at those radii were radial motion is present, since we can reasonably assume that gas is moving inwards under the gravitational torques exerted by the stellar bar and/or spiral arms (i.e. V$_{rad}$ is expected to be less or equal to 0\,km/s. Therefore, a circular velocity curve of the following form \citep[e.g.][]{1979ARA&A..17..135F} is fitted as an upper envelope to the peaks in the rotation curve (solid line in Fig. \ref{fig:rotcurve}). 
\begin{equation}\label{eq:rot}
 V_{circ}(R)=\frac{V_{max}\times(R/R_{max})}{[1/3 + 2/3\times(R/R_{max})^n]^{3n/2}}
\end{equation}
with $V_{max}$=156\,km\,s$^{-1}$, $R_{max}$=440\,pc and n=0.855. This smooth profile is indicative of the real galactic rotation profile, if all material was only rotating on circular orbits. This functional form further gives smooth curves for $\Omega$, $\Omega-\kappa/2$, etc. The difference between the fitted velocity and the model circular velocity is shown in the bottom panel of Fig. \ref{fig:rotcurve}. At radii larger than 6\,kpc there is very little deviation from the circular velocity profile. 

Inside 6\,kpc the offset reaches -\,50\,km\,s$^{-1}$. In the inner 6\,kpc the fitted and model velocity curves overlap at two radii, 1.3-1.7\,kpc and 400\,pc. To avoid resolution effects, we don't compute the difference within a conservative limit of 300\,pc. The smaller radius can be immediately linked to the radius of the outer circumnuclear ring. At $\sim$1.5\,kpc there is no ring structure, but \citet{Haan2009} made a velocity decomposition of the HI velocity field and found a strong variation in the $s3/s1$ term, which is indicative of a co-rotation (CR) radius \citep{1997ApJ...479..723C}. Also, the BIMA SONG data showed that CO(1-0) emission was constrained to this radius and in the dust spiral shown in Fig. 4a of \citet{Jogee2002} it does seem that the spiral's pitch angle changes significantly at this radius, indicating a change in gas/dust dynamics. 

The $\Omega$, $\Omega +\kappa/2$, $\Omega -\kappa/2$ and $\Omega -\kappa/4$ frequency curves, based on the fitted circular velocity curve, are also included in Fig. \ref{fig:rotcurve}. The exact shape of the frequency curves depends on the derivative of the velocity curve. It is thus very sensitive to small variations. For this reason the curves within the central kiloparsec are not drawn. Further, \citet{2004MNRAS.354..883M} shows how the inclusion of a BH changes the frequency curves without changing the velocity curve. The  $\Omega -\kappa/2$ curve may show either a downturn followed by an upturn (constant density core), or an immediate upturn (MBH or density cusp). Circumnuclear star forming rings generally form near the ILR \citep[e.g.][]{Athanassoula1992a,Regan2003,2008ApJS..174..337M, 2012ApJ...747...60K}, where this  $\Omega -\kappa/2$ curve intersects with the pattern speed of the asymmetric pattern that drives the gas inward. \citet{Jogee2002} estimate that the pattern speed for their 5.9\,kpc large scale stellar bar must be close to 30\,km/s/kpc given the semi-major axis of that bar; a bar's CR radius is typically at 1.2$\pm$0.2 times the semi-major axis of the bar \citep{Athanassoula1992a}. If the pattern speed is 30\,km/s/kpc, then the bar's CR would be at 7\,kpc and the (outer/only) ILR would be at 1.6\,kpc.

\citet{2006ApJ...644..180Y} have also derived frequency curves for NGC\,5248 by application of their nonlinear asymptotic theory of spiral density waves. Their results are based on H$\alpha$ observations of the center, the Jogee CO(1-0) data, and the BIMA SONG CO(1-0) observations and extend out to 6\,kpc. The frequency curves derived in this work lead to two ILR, with the oILR at 3\,kpc and the iILR very close to the nucleus. However, the same caveats as above should be taken into account here. 

\begin{figure}
\resizebox{\hsize}{!}{\includegraphics{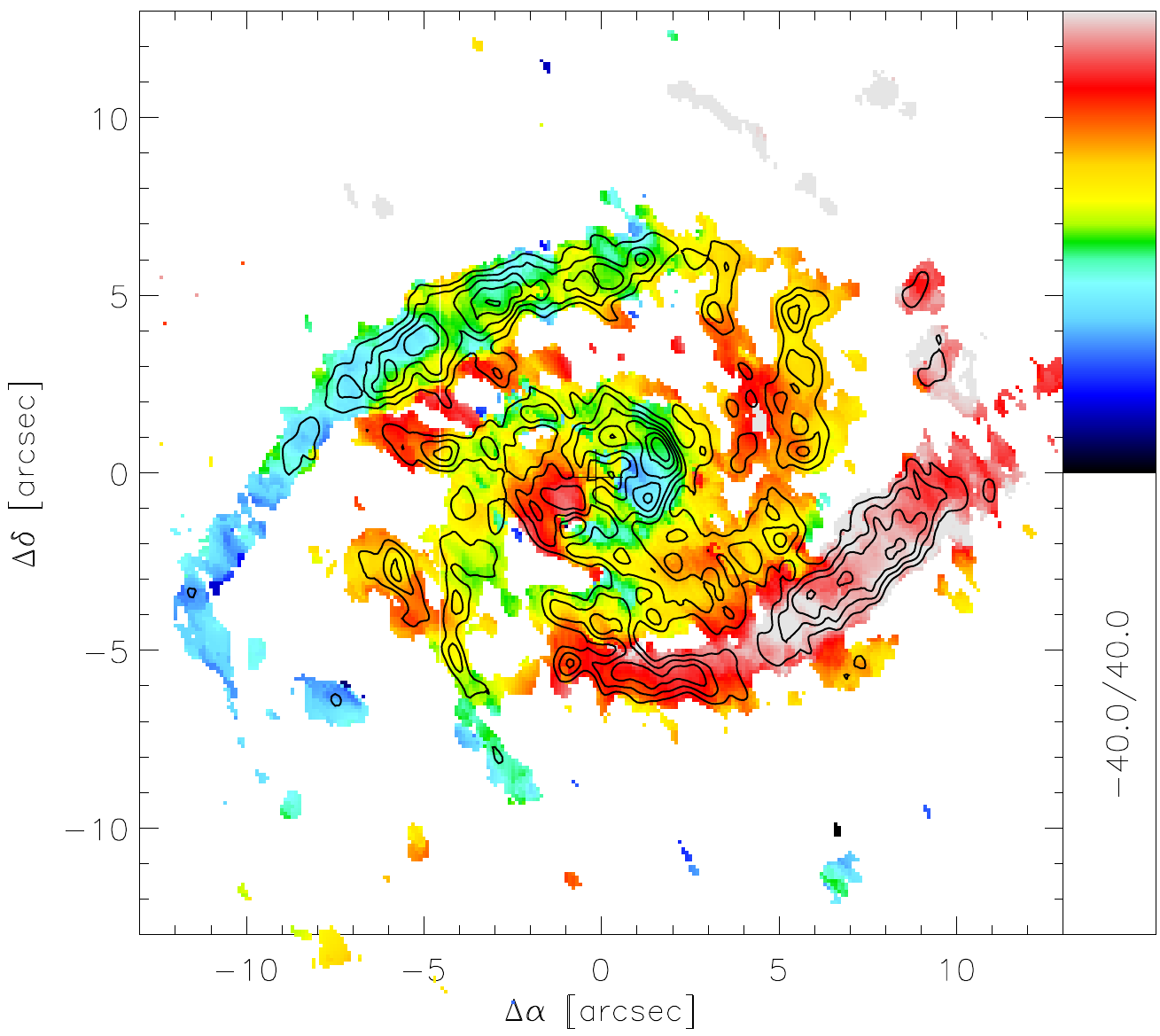}}
\caption{CO(2-1) residual velocity field, obtained by subtracting the modeled circular velocity from the observed CO(2-1) velocity field. Overlaid are the flux contours of the integrated CO(2-1) emission. Contours start at 5$\sigma$, the residual velocity field was calculated at each position where the observed velocity field exceeded a 2$\sigma$ measurement. {\it A color version of this figure is available in the electronic version.}}
\label{fig:subcirc}
\end{figure}

As a final investigation of the gas kinematic, we obtained the 2D distribution of non-circular motions in the CO(2-1) velocity field (Fig. \ref{fig:subcirc}). The residual velocity field is obtained by subtracting the fitted circular velocity, despite the caveats, from the observed velocity distribution. The residual field is dominated by the spiral arm structure (in red) in the south. This southern spiral arm shows residual velocities of up to 40\,km\,s$^{-1}$/sin({\it i})=60\,km\,s$^{-1}$. The northern spiral arm also shows non-circular motions (blue), but to a smaller extent, implying that the inflow is not symmetric along the large scale bar major axis. Similar asymmetries were present along the one-sided non-zero velocities in the minor axis pv-diagram (Fig. \ref{fig:co_pvdiagram}). In the west, connecting the outer and inner rings, the gas there also shows non-circular motions. These non-circular motions indicate that there is gas flow from the outer to the inner circumnuclear ring. Comparison with the velocity dispersion field (Fig. \ref{fig:co_kin}) shows no corresponding increase in velocity dispersions. It is not clear if this non-circular motion is connected to the northern streaming spiral arm. Finally, at the inner ring radius, we reach the resolution limit and are hesitant to over-interpret the bipolar distribution seen with positive residuals in the east and negative ones in the west, as implying outflow.

\section{Stars and star formation in the center of NGC\,5248}
In the previous section we revealed a complex pattern of radial motions that shows that gas flows inward from the outer to the inner circumnuclear ring. In this section, we derive the ages of the stellar populations in and around both rings, as well as their star formation histories, in order to further investigate the ring properties. 

\subsection{Ages of the stellar clusters in the rings}\label{sect:clusters}
\citet{Maoz} detected 507 star cluster candidates in and around the two circumnuclear star forming ring in NGC\,5248. They extracted their fluxes via aperture photometry in the following broadband filters: F220W, F336W, F547M, F814W and F160W. Here, we use a $\chi^2$ fitting routine and Starburst99 models also used in \citet{Tessel2} to re-determine the ages, masses and color excess for each star cluster in that set. The parameters of the Starburst99 models are: Starburst99 version v6.0.2 \citep{1999ApJS..123....3L,2005ApJ...621..695V,2010ApJS..189..309L}, run in burst mode, with a Kroupa IMF, solar metallicity, and Padova AGB solar metallicity tracks. Output was generated every 2\,Myr years. The new results that are presented in this work mainly reflect the improvements in stellar models over the past 10 years since the results of \citet{Maoz} were published. We also included nebular emission in the model fluxes, which adds significant emission for young ages at, predominantly, the shorter wavelengths. Our photometric library also contains a wider age range than the original \citet{Maoz} work. The generated model spectra were weighted with a range of E(B-V) values ([0,3] in 0.01 steps) and stellar masses ([10$^3$-10$^7$M$_{\sun}$], in 0.2\,dex steps). Table \ref{tab:starcluster} gives the fitted ages (column 2), color excesses (column 3) and masses (column 4) for the first 20 star clusters (the full table is available in the online appendix) that contained detections in 3 or more bands and were successfully fitted. Column 1 contains the original ID of each star cluster. In total 308 candidates were successfully fitted, $\chi^2$ given in column 5. 

\begin{table}
\begin{minipage}{\columnwidth}
\caption{Star cluster ages, extinctions and masses in the circumnuclear star forming rings of NGC\,5248}\label{tab:starcluster}
\centering
\begin{tabular}{l c c c c}
\hline\hline
ID & Age [Myr] & E(B-V) & log(Mass/M$_{\sun}$) & $\chi^2$/d.f. \\
\hline
001    &     6    &    0.00    &     4.8    & 18.29   \\
002    &    90    &    0.05    &     5.4    & 12.98   \\
003    &    16    &    0.00    &     4.8    & 10.23   \\
004    &    38    &    0.06    &     5.0    &  4.19   \\
005    &    22    &    0.38    &     5.0    & 10.87   \\
006    &    46    &    0.12    &     5.0    &  3.13   \\
007    &     6    &    0.00    &     4.0    &  6.31   \\
008    &     6    &    0.15    &     4.0    &  5.60   \\
009    &     6    &    0.20    &     4.0    &  6.93   \\
010    &    52    &    0.00    &     4.8    & 10.87   \\
011    &    28    &    0.00    &     4.6    &  7.61   \\
012    &    50    &    0.04    &     4.8    &  9.20   \\
013    &     6    &    0.05    &     3.8    &  6.54   \\
014    &     6    &    0.25    &     4.0    &  3.87   \\
015    &    42    &    0.00    &     4.6    &  4.16   \\
016    &    34    &    0.07    &     4.6    &  0.34   \\
017    &    44    &    0.00    &     4.6    &  5.41   \\
018    &    22    &    0.37    &     4.6    & 12.36   \\
019    &    26    &    0.00    &     4.4    &  3.31   \\
020    &   122    &    0.18    &     5.0    &  4.50   \\
\hline
\end{tabular}\\

\begin{flushleft}
{\bf Notes:} Results of the $\chi^2$ fitting of age (column 2), color excess (column\\ 3), and mass (column 4) of the first 20 observed star clusters of NGC\,5248. \\The star cluster ID in column 1 is equal to that in the \citet{Maoz}\\ electronic table. The full table is available in the online appendix.
\end{flushleft}
\end{minipage}
\end{table}

\begin{figure*}
\begin{tabular}{c c c}
\includegraphics[width=5.6cm]{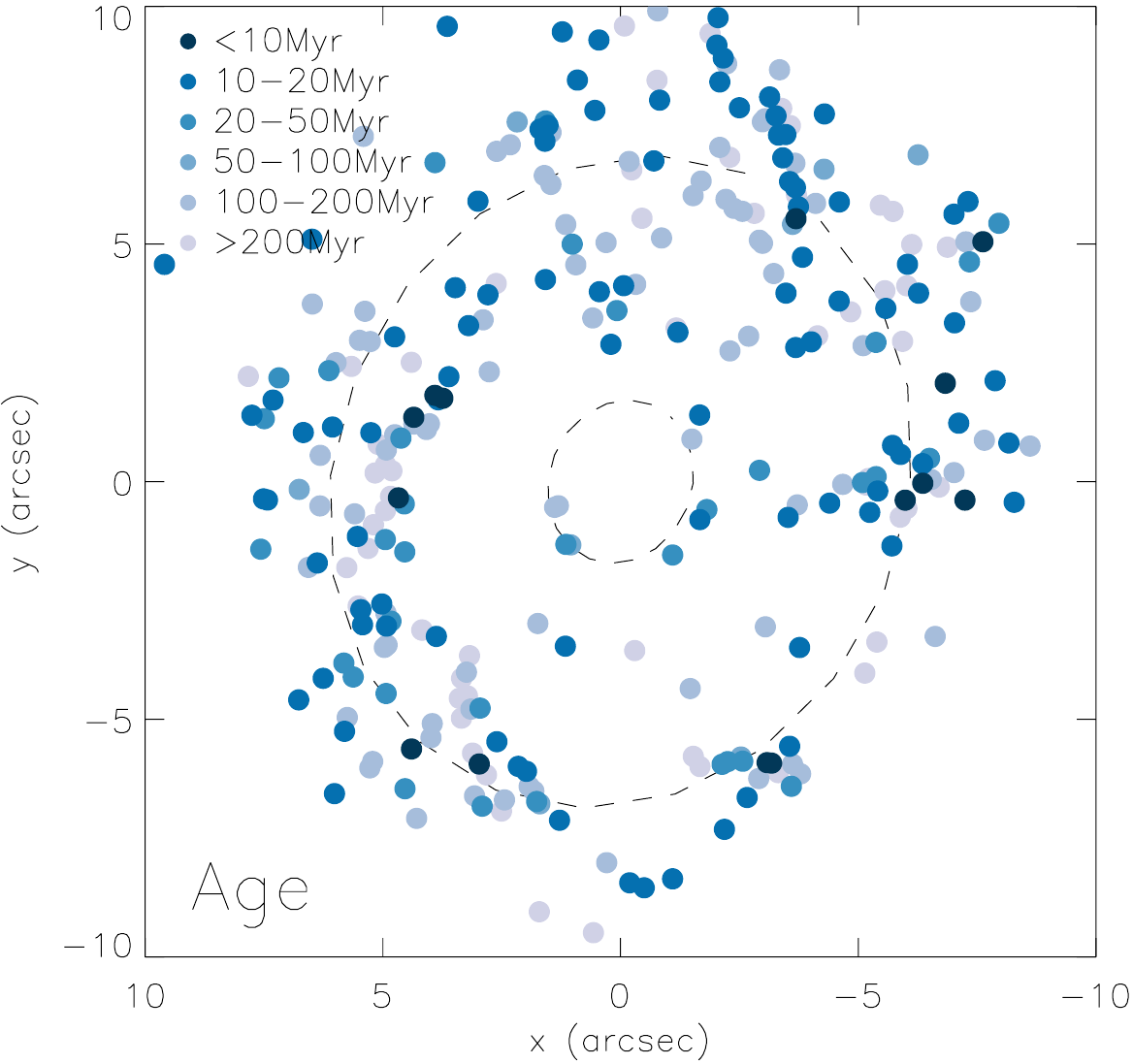}&
\includegraphics[width=5.6cm]{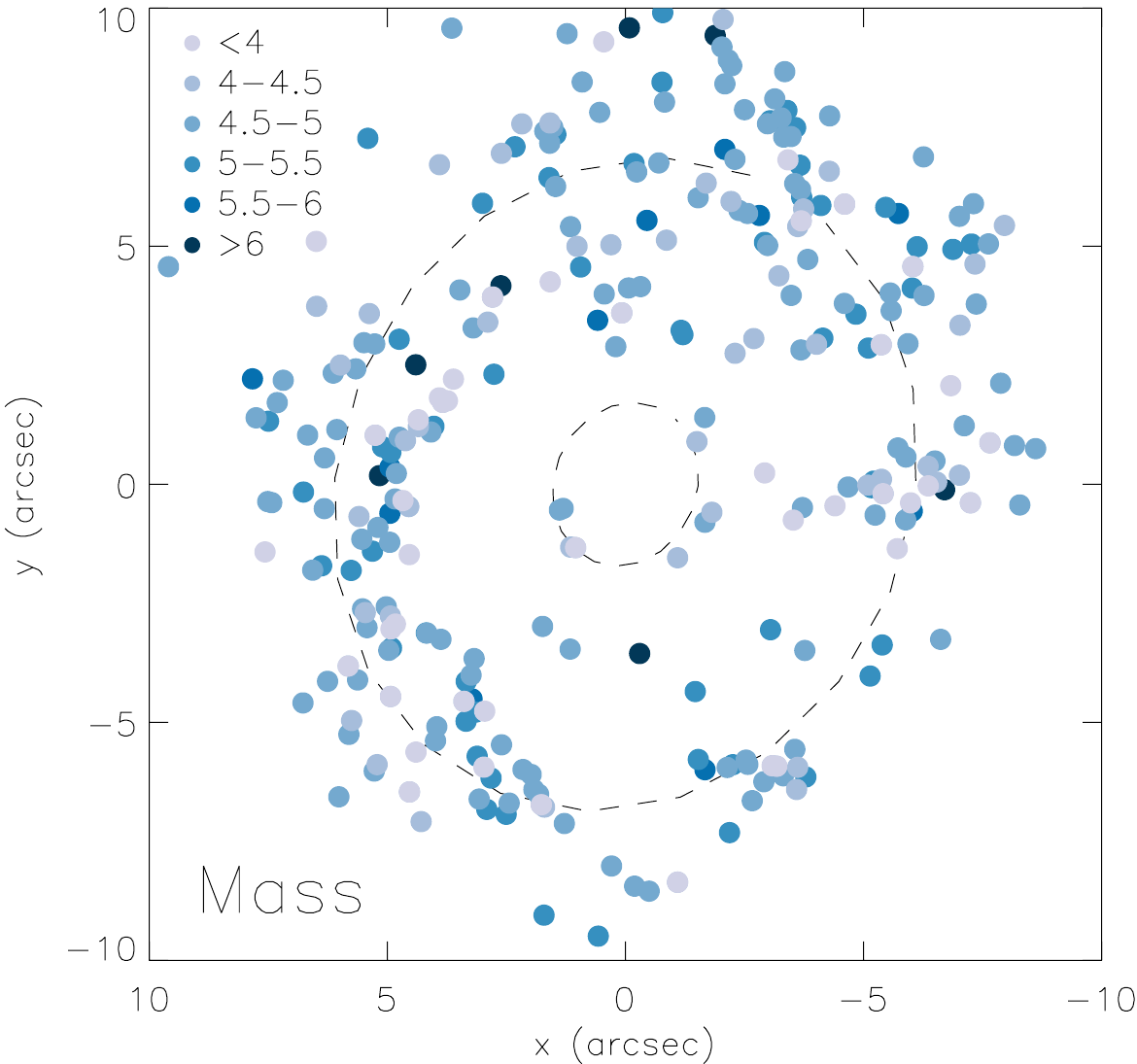}&
\includegraphics[width=5.6cm]{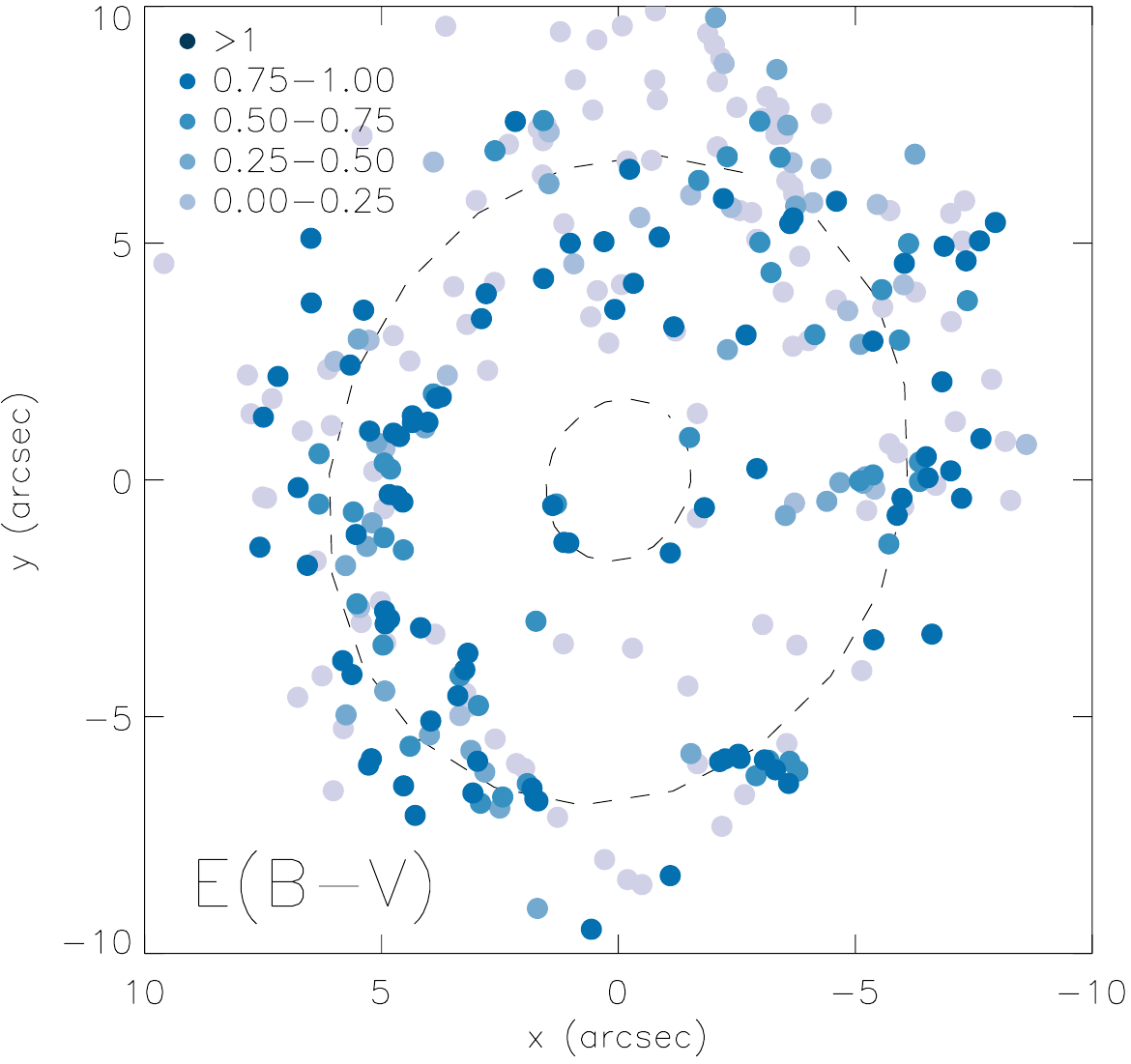}\\
\end{tabular}
\caption[Results of the $\chi^2$ fitting as function of azimuthal angle]{Results of the $\chi^2$ fitting. {\it Left} derived ages, {\it middle} derived masses, {\it right} derived color excess. The positions of the star clusters are relative to the dynamic center and have been deprojected.}
\label{fig:starclusterresults}
\end{figure*}

In comparison with previous results from \citet{Maoz}, we find that the star clusters
\begin{itemize}
\item{have slightly lower $\chi^2$ values,}
\item{are on average 30\% heavier,}
\item{are older ($>$50Myr), and}
\item{have similar color excess}
\end{itemize}
There are several factors that can explain why we obtain heavier and older star clusters. Our analysis uses a Kroupa IMF, while \citet{Maoz} use a Salpeter IMF. A Kroupa IMF has a {\it higher} fraction of low-mass stars. Most of the luminosity is contained in the high mass stars, so low mass stars can increase the total mass of a cluster without (significantly) increasing the luminosity. This result holds even when we take the difference in high mass cut-offs in the chosen IMFs into account (120$M_{\sun}$ by Maoz, 100$M_{\sun}$ by us). A second factor that changes the derived star cluster mass is that our fitting routine does not find as many young star cluster ages. The inclusion of nebular emission boosts the fluxes at the short wavelength side of the younger model SEDs and fewer star clusters fit that distribution. As a direct result, the fitted mass of the star cluster must increase, since the mass-to-light ratio increases with age. These two factors in mass `gain' are offset by a difference in the adopted distance to the galaxy. In this work a newer distance measurement of 12.7\,Mpc for NGC\,5248 \citep{2009AJ....138..323T} is adopted, while \citet{Maoz} adopted a distance of 22.7\,Mpc. The shorter distance would lower our derived cluster masses by about a factor of 3. Combined, these factors culminate in the 30\% more massive and somewhat older star clusters found in the current fitting.

In Fig. \ref{fig:starclusterresults} (left panel) the star cluster fitting output is visualized for age. The position of each star cluster is relative to the dynamic center and the positions have been deprojected. The derived ages are split into six age ranges, the color hue of each bin decreases with age. The youngest ($<$10\,Myr) star clusters are found in the larger ring, in its west and east quadrants. Older star clusters are distributed uniformly throughout the larger ring. At the inner ring, only seven star clusters were successfully fitted. There, the ages of the clusters start in the 10-20\,Myr age range and increase along a clockwise gradient from the north. The derived masses of the star clusters are shown in the middle panel. The results are binned by 0.5 dex. The star clusters have an average stellar mass of 10$^{4.5}M_{\sun}$. No significant mass variation as a function of azimuth can be seen in either ring. Finally, the derived color excesses are plotted in Fig. \ref{fig:starclusterresults} (right panel). The color hue increases with increasing color excess. The color excess is mostly uniformly distributed throughout the rings. Star clusters with lower color excess are primarily found in the north and at the outer edge of the larger ring. At the inner ring again a clockwise gradient is seen, with increasing color excess from the north. 

A comparison of the age and color excess azimuthal gradient in the smaller ring with the strongest emission of the CO(2-1) line shows that the highest extinction is partly co-spatial with the CO(2-1) emitting gas. The youngest star clusters in this inner circumnuclear ring are positioned at a slightly larger clock-wise angle. Therefore, star formation appears predominantly to occur at one location. No [FeII] emission was detected from the inner ring \citep{Boker2008}. [FeII] emission is believed to trace supernova activity, which starts $\geq$10\,Myr after star formation. The lack of [FeII] emission, combined with the age gradient of star clusters found in this ring, suggests that star formation is intermittent, possibly periodic, in the inner ring. In Sect. \ref{sect:nestedbar} we speculate how this gradient of extinction, gas and young stars could be generated.

\begin{figure*}
\centering
\includegraphics[width=17cm]{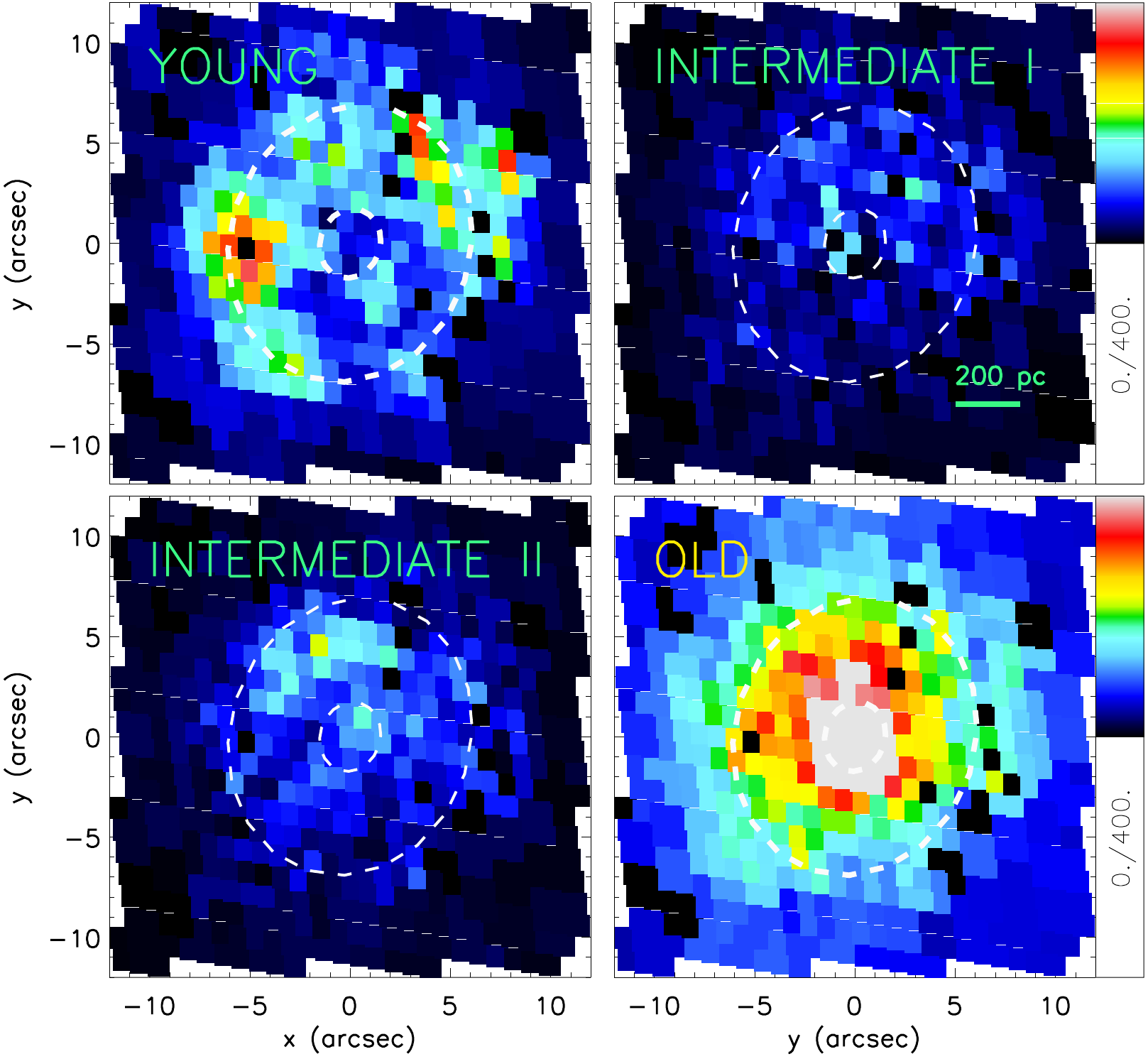}
\caption{Stellar `mass' maps in [a.u.] for the 4 age bins fitted to the \Sauron\, IFU data set. The maps have been corrected for inclination and the positions of the two circumnuclear rings (1.5\arcsec\, and 6\arcsec\,) have been indicated. {\it A color version of this figure is available in the electronic version.}}
\label{fig:IFU_SSP}
\end{figure*}

\subsection{Stellar populations from SAURON observations}\label{sect:IFU}
The star cluster analysis above is supplemented with an analysis of the underlying stellar population as traced by \Sauron\, IFU observations. The flux at each \Sauron\, spaxel contains the contributions of all stellar populations at those positions, including stars that are not in any of the detected star clusters. The \Sauron\, data cube was originally fitted with a full range of age and metallicity SSP models from the MILES database\footnote{http://miles.iac.es}. The resulting maps are shown in Fig. \ref{fig:sauron}. This large fit shows that solar metallicity is the predominant metallicity in the central region of this galaxy. A fitting residual of $\sim$1.5\% of the flux remains in each spaxel when fitting with this full range of SSP models. 

Following \citet{Tessel2}, a smaller combination of SSP spectra, at solar metallicity, is fitted to the \Sauron\, data cube. The goal is to separate the stellar light at each spaxel into `young', `intermediate' and `old' components, and determine if the rings stand out in recent star formation. For this reason, four SSP model spectra are selected. The four SSPs chosen have solar metallicity and have designated model ages of 70.8\,Myr, 158.5\,Myr, 398.1\,Myr and 3.2\,Gyr. Van der Laan et al. determined that these four SSP spectra reasonably span the full possible stellar age range. The emission lines, H$\beta$, [OIII], and [NI] are also included in this alternative fitting to avoid contamination of the SSP results by gaseous emission. 

The output from the fitting routines pPXF \citep{CappEms04, Emsellem+04} and GANDALF \citep{Sarzi+06} contains the number weighted fractions of each SSP spectrum at each spaxel. From these number weighted fractions the {\it mass} weighted fractions at each position are determined, via the luminosity of each SSP spectrum and its mass-to-light ratio. This mass fraction is multiplied with the stellar intensity at each position to give an `absolute' mass distribution in each SSP age bin. The age bins are designated `young', `intermediate I', `intermediate II' and `old'.  We iterate the fitting 30 times. Random noise on the level of the initial fitting residual (1.5\% of the flux) is introduced at each iteration. In this manner, a mass fraction uncertainty of less than 5\% can be reached. The results are presented in Fig. \ref{fig:IFU_SSP}.

In the `young' age bin, high stellar masses are predominantly seen at two locations opposite each other at the radius of the outer ring. The majority of the young star clusters are also found at these locations. At the inner ring some lower fractions of young stars are seen.

The `intermediate' age bins show the least stellar mass of all panels. \citet{Tessel2} investigated this. The fitting routine minimizes the number of SSP spectra needed to achieve an acceptable fit. Therefore, the high(est) contrast between `young' and `old' is used when possible. Nevertheless, some intermediate age stellar populations are found at both the outer and inner circumnuclear ring. The `intermediate II' bin shows a significant fraction of stellar mass in the NE quadrant of the outer ring. This is between the two locations of high `young' stellar mass and co-spatial with one of the high 1\,mm emission peaks as seen in Fig.\ref{fig:1mmcont}. Neither ring is visible as such in the `old' stellar mass map. The `old' distribution shows a `bulge-like' profile, peaking in the center. 

\section{Discussion}
In understanding the make-up of the central region in NGC\,5248 we need to account for both the gaseous and stellar distribution, as well as the star formation that connects then. To this end we first discuss the locations were star formation occurs. Afterwards, we discuss which components are a likely part of the central region of this galaxy. 

\subsection{Star formation scenarios in the circumnuclear rings}\label{sect:agegradient}
When gas is driven onto a circumnuclear ring due to driving by a stellar bar, two star formation scenarios in circumnuclear rings are proposed. They are called `popcorn' and `pearls-on-a-string'. The difference between the two lies in the assumed time between gas inflow and the onset of star formation. The `popcorn' scenario was proposed by \citet{Elmegreen1994} and holds that the gas density in the ring builds up throughout the ring to some critical density, at which point star formation starts uniformly throughout the ring. The second scenario was put forward by \citet{Boker2008} and assumes that star formation will occur predominantly close to the two `inflow' or overdensity points of gas onto the ring. The argument is that the gas density will be higher where the gas spiral arms and ring connect than elsewhere in the ring and thus that star formation will mainly occur there. As the new stars are moving slowly out of the overdensity regions, a string of aging stellar populations would be formed in the ring. Evidence for the `pearls-on-a-string' scenario in one or both rings is thus an indication for the origin of that ring to be due to gravitational torques from a stellar bar.

In both the star cluster analysis and the \Sauron\, fitting, the youngest stellar populations are found at two locations. These positions are north-west and south-east at the radius of the larger ring. They are anti-correlated with the CO emission at this radius, which could imply that molecular gas is almost fully converted into stars as it enters the ring, the 'pearls-on-a-string' scenario. Arguments for `pearls' may further be found in the 1\,mm continuum map (Fig. \ref{fig:1mmcont}). Two high emission regions are present in the 1\,mm continuum distribution at the radius of the larger circumnuclear star forming ring. 1\,mm emission arises most likely from cold dust and traces the bulk of the gas mass in the circumnuclear region. The combination of the high emission regions in the 1\,mm continuum map/CO spiral arms and the young stellar mass concentration would form two `pearls' of the `pearls-on-a-string' scenario; current star formation and recent star formation. The contact points of the spiral arms generated by the Jogee et al. 5.9\,kpc large scale stellar bar (PA=135\degr\,) should have gas entering the larger circumnuclear ring at the north and south. This corresponds with our observations only if star formation is somewhat delayed (a quarter rotation, $\sim$5\,Myr).

There are some indications that the star formation in the inner ring also shows sites of preferred star formation. The fitted star clusters are found exclusively in the western-southern half of the ring. A sequence was discussed in Sect. \ref{sect:clusters}, going from position with the highest extinction (which could be interpreted as an overdensity point), followed by/partly co-spatial with the CO(2-1) emitting gas, followed by the location of the youngest star clusters in this inner circumnuclear ring. In analogy with the outer circumnuclear ring, there would be 3 `pearls', but starting only at one contact point! 

It is not possible to obtain a more quantitative observation of the `pearls-on-a-string' scenario. Orbital timescales in the central kiloparsec are $\frac{\pi}{v\,{\rm [km/s]}/ r{\rm \,[kpc]}} \sim$5-20\,Myr. At the same time, a reasonable timescale for a large scale bar pattern to complete one rotation, which sets the rotation of the overdensity points, is $\sim$100\,Myr. In one rotation of the pattern, the newly formed stars would have made several full rotations already. This makes detecting any stellar age gradient beyond the very crude and broad ones just discussed, impossible. However, we should also not overweight the `pearl' observation, especially at the inner ring. Since only one contact point is found, and the dynamical times are truly very short, it is still possible that this is a transient configuration.

\subsection{What components drive the central region?}\label{sect:nestedbar}
Gravitational mechanisms dominate gas transport on kiloparsec scales \citep{2005A&A...441.1011G,Haan2009}, so our first focus is on identifying the dominant mass component. The gravitational potential in the central 1.5\,kpc in NGC\,5248 is dominated by the stellar mass, as can be concluded from the following simple calculation. The dynamical mass M$_{dyn}$ is $\sim$\,8.5$\times$10$^9$\,M$_{\sun}$ at 1.5\,kpc radius, of which the atomic (HI) and molecular (CO) gas together make up ~10\%. This implies that gas self-gravity can be largely ignored and that the (molecular) gas distribution is shaped by the stellar gravitational potential. 

\citet{Jogee2002} established the current understanding of this system with their discovery of a 95\arcsec\, (5.9\,kpc) stellar bar. This `weak' (ellipticity 0.44) oval/bar drives two spiral arms inside its CR. These arms show star formation between 30\arcsec\, and 90\arcsec\,.  Since star formation requires that gas is not `streaming', it is unlikely that there is significant radial gas flow inwards at these radii. However, this spiral arm/bar system also is visible in dust maps at radii between 6\arcsec\, and 70\arcsec\,.  Between 6\arcsec\, and 30\arcsec\, there is no star formation in these spiral arms. \citet{Jogee2002} argue that the large-scale stellar bar drives an \citet{Englmaier2000}-type density wave spiral that crosses the ILR of the bar pattern at 26\arcsec\, (1.6\,kpc) and accounts for the full distribution of the molecular gas in this system. 

In the case of this scenario, we have an objection to the extent of the nuclear spiral. \citet{Jogee2002} extend the density wave spiral down to the smaller circumnuclear ring at 1.5\arcsec\, (100\,pc). This argument is based on two observations. \citet{1999MNRAS.302L..33L} detected a grand-design nuclear spiral in NIR J-, H-, and K-band observations. However, the luminosity contrast between arm and inter-arm regions is only 0.05\,mag. We therefore hold it to be tentative. The second indicator is the existence of a `patchy' spiral arm seen in a K$_s$-band \citep{Jogee2002} observation of the region between the two circumnuclear rings, that could be connected to one of the spiral arms at larger radii. The presence of a second arm, that connects to the other spiral, is not clear from those observations. \citet{2006ApJ...644..180Y} also determined the form of the spiral density wave and claimed its existence to 100\,pc. Those authors acknowledge that, possibly due to loss of resolution, the spiral merges into an oval feature at these radii. 

We do find non-circular motion between the two rings in our CO(2-1) observations. Indeed, such motion is necessary to fuel the current star formation at the smaller circumnuclear ring. However, we find no clear indication in our data that the molecular gas is indeed still driven by the density spiral wave inside the 5\arcsec\, (370\,pc) star forming ring. Specifically, because it is a star forming ring with clear indications that gas is rapidly being converted into stars, the `pearls-on-a-string' scenario discussed in the previous section.

An alternative scenario may be the following. A dynamical analysis by \citet{Haan2009} found 2 CR in NGC\,5248, at 28-30\arcsec\, (1.7-2.0\,kpc) and 120-185\arcsec\, (7.5-12\,kpc), indicating two patterns. The latter may indeed refer to the large scale stellar bar/spiral arms, but the first is still to be explained. The 28-30\arcsec\, radius is similar to the star formation transition, as well as the BIMA SONG CO(1-0) molecular gas limit of 22\arcsec\, (1.4\,kpc). 

Earlier estimates \citep[e.g.][]{1995AJ....109.2428M} of a stellar bar in NGC\,5248 were of a smaller bar, only 22\arcsec\, (1.4\,kpc) in length, but with similar PA to the 95\arcsec\, (5.9\,kpc) bar. There are several arguments to reconsider the inclusion of this smaller bar into our understanding of this system. It explains the second CR. Also, the transition at 30\arcsec\, from non star forming to star forming spiral arms is explained. Inside the CR of this smaller bar, the spiral arms would indeed be dominated by shocks, thus non-star forming and given to streaming motion. Further, it is much more probably that the 6\arcsec\, (370\,pc) circumnuclear ring is near the (i)ILR of this smaller stellar bar. 

It is our belief that the formation of the smaller circumnuclear ring at 100\,pc remains undetermined. Given the strong star formation at the larger ring, it is very possible that the increased radiation pressure and/or turbulence of the young stars leads to gas overcoming the barrier and moving further in. These hydrodynamical forces especially play a role when a significant radial density gradient is present, as in a circumnuclear ring. The bimodal structure of the star formation in the larger ring could also explain the \citet{1999MNRAS.302L..33L} grand-design nuclear spiral observation. Finally, due to the presence of an SMBH at the center, a nuclear resonance would halt the inflowing gas again before reaching the nucleus proper. 

Either scenario would be consistent with the observed stellar and gaseous distribution in the central region of this galaxy. 

\section{Summary}
The central region of the barred spiral galaxy NGC\,5248 contains two circumnuclear star forming rings at 1.5\arcsec\, (100\,pc) and 6\arcsec\, (370\,pc) from its quiescent nucleus. This distribution is surprising, since circumnuclear star forming rings are considered to be efficient gas barriers. We combine analyses of the gaseous and stellar content in the central kiloparsec of this galaxy to understand the gas distribution and dynamics of this star forming central region. We present new PdBI+30m CO(2-1) emission line observations that show two spiral arms at the larger ring radius, and a gas ring at the smaller ring radius. This is the first time that the 100\,pc circumnuclear ring is resolved in molecular gas observations.

From the stellar analysis it has become clear that the 6\arcsec\, (370\,pc) star forming ring is a `true' circumnuclear ring. We find evidence of the `pearls-on-a-string' star formation scenario, which implies that gas driven onto this ring is quickly converted into stars. Further, we find that there is radial gas motion between the two rings, consistent with current star formation at the inner ring. 

The current understanding of this region comes from \citet{Jogee2002}. They propose that the gas distribution is dominated by a spiral density wave driven by a large scale stellar bar of 95\arcsec\, (5.9\,kpc). Following \citet{Englmaier2000} this density wave crosses the ILR at 26\arcsec\, (1.6\,kpc) radius to become a nuclear spiral. This nuclear spiral is believed to continue down to 1.5\arcsec\, (100\,pc) and thus fuel both star forming rings. 

We propose an alternative scenario for consideration. Resurrection of a stellar bar previously believed to be present \citep{1995AJ....109.2428M}, which has an extent to 22\arcsec\, (1.4\,kpc), in combination with the larger stellar bar/spiral arms. This stellar bar would explain the two CR found by \citet{Haan2009}, as well as provide an alternative explanation for the transition from not star forming to star forming at 30\arcsec\, (1.8\,kpc) radius in the spiral arms. Furthermore, a smaller bar is better placed to explain the location of the 6\arcsec\, (370\,pc) star forming ring.

For the origin of the 1.5\arcsec\, (100\,pc) star forming ring, we believe there is not enough evidence to conclude a nuclear spiral wave. Viscous forces at the larger ring are equally probable to have brought gas further inward. The resonance near which this smaller ring forms is most likely a nuclear resonance with the SMBH.

\begin{acknowledgements}
We thank the referee for his/her comments that helped improve this paper. TvdL was in part supported by DFG-funding (grant SCHI 536/2-3). This work is based on observations carried out with the IRAM Plateau de Bure Interferometer. IRAM is supported by INSU/CNRS (France), MPG (Germany) and IGN (Spain). We would like to thank the staff at IRAM Spain, especially A. Sievers and M. Gonzalez, for invaluable help with the 30m observations, and our local contact for the PdBI at IRAM France, J.M. Winters. We also thanks D. Maoz for making his {\it HST} observations available to us. This research has made use of the NASA/IPAC Extragalactic Database (NED) which is operated by the Jet Propulsion Laboratory, California Institute of Technology, under contract with the National Aeronautics and Space Administration.
\end{acknowledgements}

\bibliographystyle{aa}
\bibliography{papers.bib}

\Online
\begin{appendix}
\onecolumn
\section{Table \ref{tab:starcluster} -- full}
\begin{longtable}{l c c c c}
\caption{\label{fulltable} Star cluster ages, extinctions and masses in the circumnuclear star forming rings of NGC\,5248} \\
\hline\hline
\multicolumn{1}{c}{ID} & Age [Myr] & E(B-V) & log(Mass/M$_{\sun}$) & $\chi^2$/d.f. \\ \hline
\endfirsthead
\multicolumn{5}{c}%
{\tablename\ \thetable{} -- continued.} \\
\hline \hline \multicolumn{1}{c}{ID} & Age [Myr] & E(B-V) & log(Mass/M$_{\sun}$) & $\chi^2$/d.f. \\ \hline
\endhead
\hline
\endfoot
001    &     6    &    0.00    &     4.8    & 18.29   \\
002    &    90    &    0.05    &     5.4    & 12.98   \\
003    &    16    &    0.00    &     4.8    & 10.23   \\
004    &    38    &    0.06    &     5.0    &  4.19   \\
005    &    22    &    0.38    &     5.0    & 10.87   \\
006    &    46    &    0.12    &     5.0    &  3.13   \\
007    &     6    &    0.00    &     4.0    &  6.31   \\
008    &     6    &    0.15    &     4.0    &  5.60   \\
009    &     6    &    0.20    &     4.0    &  6.93   \\
010    &    52    &    0.00    &     4.8    & 10.87   \\
011    &    28    &    0.00    &     4.6    &  7.61   \\
012    &    50    &    0.04    &     4.8    &  9.20   \\
013    &     6    &    0.05    &     3.8    &  6.54   \\
014    &     6    &    0.25    &     4.0    &  3.87   \\
015    &    42    &    0.00    &     4.6    &  4.16   \\
016    &    34    &    0.07    &     4.6    &  0.34   \\
017    &    44    &    0.00    &     4.6    &  5.41   \\
018    &    22    &    0.37    &     4.6    & 12.36   \\
019    &    26    &    0.00    &     4.4    &  3.31   \\
020    &   122    &    0.18    &     5.0    &  4.50   \\
021    &    28    &    0.01    &     4.4    & 11.30   \\
022    &    34    &    0.27    &     4.6    &  1.57   \\
023    &    38    &    0.22    &     4.6    &  1.26   \\
024    &   122    &    0.23    &     5.0    &  5.19   \\
025    &     6    &    0.36    &     3.8    & 10.19   \\
026    &    12    &    0.00    &     4.0    &  3.01   \\
029    &    36    &    0.00    &     4.4    &  4.56   \\
030    &   122    &    1.00    &     5.4    & 16.86   \\
032    &   122    &    0.56    &     5.2    & 13.40   \\
033    &   122    &    0.11    &     4.8    &  2.79   \\
034    &    30    &    0.17    &     4.4    &  2.90   \\
036    &   220    &    0.51    &     5.2    &  0.52   \\
037    &   998    &    0.42    &     5.6    &  4.07   \\
038    &    20    &    0.00    &     4.2    &  1.03   \\
039    &   122    &    0.13    &     4.8    &  3.47   \\
040    &    14    &    0.00    &     4.0    &  0.76   \\
041    &   122    &    0.49    &     5.0    &  3.22   \\
042    &    20    &    0.03    &     4.2    &  6.11   \\
043    &   122    &    0.47    &     5.0    &  3.09   \\
044    &     6    &    0.16    &     3.6    &  0.32   \\
045    &    42    &    0.00    &     4.4    &  1.15   \\
046    &   150    &    0.12    &     4.8    &  0.52   \\
048    &    14    &    0.01    &     4.0    &  0.79   \\
052    &   850    &    1.86    &     6.4    & 45.91   \\
053    &    16    &    0.27    &     4.2    &  5.03   \\
054    &   124    &    0.31    &     4.8    &  0.08  \\
055    &    26    &    0.00    &     4.2    &  1.90   \\
056    &    52    &    0.09    &     4.4    &  3.33   \\
057    &   122    &    0.14    &     4.8    &  6.24   \\
058    &   132    &    1.94    &     5.8    &  6.75   \\
059    &    16    &    0.00    &     4.0    &  4.91   \\
060    &     6    &    0.38    &     3.6    &  2.49   \\
062    &   212    &    0.09    &     4.8    &  0.34   \\
063    &    30    &    0.35    &     4.4    &  1.49   \\
064    &    16    &    0.69    &     4.4    &  3.71   \\
065    &   122    &    0.31    &     4.8    &  0.31   \\
066    &     6    &    0.14    &     3.4    &  1.73   \\
067    &    30    &    0.00    &     4.2    &  4.40   \\
068    &   122    &    0.40    &     4.8    &  4.25   \\
069    &   132    &    0.06    &     4.6    &  0.14  \\
070    &   122    &    0.03    &     4.6    &  2.74   \\
071    &   112    &    0.00    &     4.6    &  2.49   \\
072    &    36    &    0.01    &     4.2    &  0.42    \\
073    &    30    &    0.02    &     4.2    &  0.18   \\
074    &    22    &    0.23    &     4.2    &  3.69   \\
075    &   122    &    0.41    &     4.8    &  1.05   \\
078    &   190    &    0.21    &     4.8    &  0.19   \\
079    &    16    &    0.12    &     4.0    &  0.76   \\
081    &    44    &    0.00    &     4.2    &  0.31   \\
082    &   246    &    1.03    &     5.4    &  2.38   \\
083    &     6    &    0.14    &     3.4    &  1.78   \\
084    &   122    &    0.06    &     4.6    &  1.67   \\
087    &   122    &    0.07    &     4.6    &  4.25   \\
088    &   122    &    0.76    &     5.0    &  5.81   \\
089    &   122    &    0.45    &     4.8    &  1.73   \\
090    &   122    &    1.01    &     5.2    & 18.45   \\
091    &   136    &    1.12    &     5.2    &  5.07   \\
094    &    32    &    0.48    &     4.4    &  1.22   \\
095    &   122    &    0.13    &     4.6    &  0.31   \\
096    &   122    &    1.39    &     5.4    & 16.53   \\
097    &   132    &    0.47    &     4.8    &  0.75   \\
098    &   132    &    0.17    &     4.6    &  0.63   \\
099    &   132    &    1.12    &     5.2    &  8.68   \\
100    &   122    &    0.79    &     5.0    &  3.64   \\
102    &   850    &    0.75    &     5.6    &  5.65   \\
103    &   152    &    0.08    &     4.6    &  1.80   \\
104    &    16    &    1.78    &     5.0    &  7.55   \\
106    &   850    &    0.81    &     5.6    &  9.51   \\
107    &   122    &    1.13    &     5.2    &  0.44   \\
108    &    16    &    0.54    &     4.2    & 16.84   \\
110    &   140    &    0.18    &     4.6    &  0.54   \\
111    &    16    &    0.49    &     4.2    &  3.73   \\
112    &    54    &    0.00    &     4.2    &  3.11   \\
113    &     6    &    0.18    &     3.4    &  1.23   \\
114    &    28    &    0.32    &     4.2    &  2.64   \\
115    &   544    &    0.40    &     5.2    &  2.17   \\
116    &   122    &    0.52    &     4.8    &  8.81   \\
117    &   122    &    0.79    &     5.0    &  3.87   \\
118    &     6    &    0.03    &     3.2    &  1.95   \\
119    &    16    &    1.38    &     4.8    &  2.53   \\
120    &    24    &    0.00    &     4.0    &  0.08   \\
121    &    22    &    0.09    &     4.0    &  2.44   \\
122    &   104    &    0.00    &     4.4    &  6.49   \\
123    &   122    &    0.57    &     4.8    &  3.44   \\
125    &   124    &    0.60    &     4.8    &  0.90   \\
126    &    16    &    0.21    &     4.0    &  0.88   \\
127    &   122    &    0.25    &     4.6    &  0.92   \\
128    &    16    &    0.21    &     4.0    &  6.21   \\
129    &   120    &    0.00    &     4.4    &  0.30   \\
130    &   136    &    0.94    &     5.0    &  3.48   \\
131    &   262    &    0.00    &     4.6    &  2.35   \\
132    &   124    &    0.63    &     4.8    &  4.42   \\
133    &   102    &    0.07    &     4.4    &  0.79   \\
135    &   124    &    0.61    &     4.8    &  2.62   \\
136    &   280    &    0.56    &     5.0    &  0.38   \\
137    &   114    &    0.00    &     4.4    &  3.70   \\
139    &   172    &    0.79    &     5.0    &  0.38   \\
140    &   106    &    0.38    &     4.6    &  0.25   \\
141    &   484    &    0.57    &     5.2    &  0.76   \\
142    &   264    &    0.59    &     5.0    &  0.41   \\
143    &    16    &    1.34    &     4.6    &  7.60   \\
146    &   122    &    0.33    &     4.6    &  1.86   \\
147    &   202    &    0.47    &     4.8    &  2.82   \\
149    &   122    &    0.00    &     4.4    &  0.74   \\
150    &   122    &    1.54    &     5.4    & 12.93   \\
151    &   850    &    0.59    &     5.4    &  2.54   \\
155    &    34    &    0.00    &     4.0    &  0.10  \\
156    &    92    &    0.00    &     4.2    &  0.38   \\
157    &   850    &    1.27    &     5.8    &  9.61   \\
158    &   802    &    0.00    &     5.0    &  1.42   \\
159    &    16    &    1.34    &     4.6    &  4.08   \\
160    &   122    &    0.07    &     4.4    &  0.59   \\
161    &   202    &    0.46    &     4.8    &  0.68   \\
162    &    16    &    1.31    &     4.6    &  6.06   \\
163    &   836    &    0.31    &     5.2    &  1.78   \\
165    &   850    &    0.72    &     5.4    &  2.99   \\
167    &   850    &    0.71    &     5.4    &  4.03   \\
168    &   132    &    1.06    &     5.0    &  3.49   \\
171    &   850    &    0.07    &     5.0    &  2.77   \\
172    &   122    &    1.05    &     5.0    & 13.68   \\
175    &   850    &    1.32    &     5.8    & 19.25   \\
176    &   202    &    0.20    &     4.6    &  0.34   \\
177    &   626    &    0.00    &     4.8    &  0.31   \\
179    &   226    &    0.46    &     4.8    &  0.33   \\
180    &    34    &    0.12    &     4.0    &  1.32   \\
182    &   104    &    0.26    &     4.4    &  1.15   \\
183    &    68    &    0.74    &     4.6    &  0.09    \\
184    &   948    &    0.73    &     5.4    &  3.50   \\
185    &   122    &    0.65    &     4.8    &  8.75   \\
186    &   850    &    1.07    &     5.6    & 20.19   \\
187    &    38    &    0.02    &     4.0    &  0.21    \\
188    &   972    &    0.10    &     5.0    &  0.29   \\
189    &   334    &    0.64    &     5.0    &  0.53   \\
190    &   850    &    0.31    &     5.2    &  8.27   \\
191    &   100    &    0.00    &     4.2    &  0.27   \\
193    &    38    &    0.05    &     4.0    &  0.11    \\
194    &    72    &    0.15    &     4.2    &  0.02    \\
195    &   122    &    0.14    &     4.4    &  0.24   \\
197    &   122    &    0.47    &     4.6    &  1.14   \\
198    &    18    &    0.08    &     3.8    &  0.24   \\
199    &   668    &    0.02    &     4.8    &  0.57   \\
200    &   122    &    1.13    &     5.0    &  5.03   \\
201    &   850    &    1.73    &     6.0    & 15.51   \\
203    &   156    &    0.77    &     4.8    &  2.43   \\
204    &   850    &    0.16    &     5.0    &  1.63   \\
207    &   122    &    0.20    &     4.4    &  0.21   \\
208    &   132    &    1.17    &     5.0    &  4.33   \\
210    &    36    &    0.17    &     4.0    &  0.41   \\
211    &   238    &    0.53    &     4.8    &  0.44   \\
213    &    16    &    0.37    &     4.0    &  4.24   \\
214    &    22    &    1.33    &     4.6    &  1.81   \\
216    &   136    &    1.49    &     5.2    &  4.25   \\
217    &   998    &    0.78    &     5.4    &  0.70   \\
218    &    16    &    0.79    &     4.2    &  6.47   \\
219    &    16    &    1.48    &     4.6    &  5.83   \\
220    &    16    &    0.88    &     4.2    &  2.49   \\
221    &   132    &    0.89    &     4.8    &  3.13   \\
222    &    16    &    1.57    &     4.6    &  3.12   \\
223    &   850    &    2.11    &     6.2    &  8.27   \\
224    &   126    &    0.88    &     4.8    &  2.75   \\
225    &   378    &    0.65    &     5.0    &  0.50   \\
226    &   850    &    0.22    &     5.0    &  0.69   \\
227    &   850    &    0.50    &     5.2    &  2.78   \\
229    &    16    &    2.17    &     5.0    &  3.36   \\
231    &   482    &    0.84    &     5.2    &  0.18   \\
232    &   124    &    1.84    &     5.4    & 19.71   \\
233    &   948    &    0.55    &     5.2    &  2.02   \\
234    &   136    &    1.20    &     5.0    &  4.05   \\
237    &   668    &    0.03    &     4.8    &  1.69   \\
239    &   122    &    0.94    &     4.8    &  1.07   \\
241    &   104    &    0.07    &     4.2    &  0.26   \\
242    &   122    &    0.26    &     4.4    &  0.73   \\
243    &   122    &    1.53    &     5.2    & 25.08   \\
244    &   322    &    0.20    &     4.6    &  0.05    \\
245    &   850    &    0.64    &     5.2    &  2.57   \\
247    &    26    &    0.00    &     3.8    &  0.27   \\
248    &   126    &    1.61    &     5.2    & 16.14   \\
250    &   968    &    0.30    &     5.0    &  0.37   \\
251    &   354    &    0.47    &     4.8    &  0.15   \\
253    &    32    &    0.00    &     3.8    &  0.53   \\
254    &   116    &    0.09    &     4.2    &  0.21   \\
256    &   122    &    0.56    &     4.6    &  1.71   \\
257    &   122    &    1.26    &     5.0    &  9.58   \\
259    &   122    &    0.07    &     4.2    &  0.10   \\
261    &   850    &    0.34    &     5.0    &  1.93   \\
262    &   140    &    0.96    &     4.8    &  1.09   \\
263    &    20    &    0.29    &     3.8    &  0.78   \\
264    &   288    &    0.30    &     4.6    &  0.03   \\
265    &   122    &    1.23    &     5.0    &  5.20   \\
266    &   850    &    0.99    &     5.4    &  5.22   \\
268    &   186    &    0.48    &     4.6    &  0.52   \\
269    &     6    &    0.00    &     3.0    &  2.76   \\
270    &   140    &    0.05    &     4.2    &  0.05   \\
271    &   122    &    0.08    &     4.2    &  1.20   \\
272    &   132    &    0.38    &     4.4    &  0.01  \\
273    &    16    &    2.44    &     5.2    &  3.03   \\
275    &    40    &    0.00    &     3.8    &  0.04   \\
276    &   292    &    0.62    &     4.8    &  0.18   \\
277    &   122    &    1.66    &     5.2    & 13.95   \\
278    &   850    &    1.65    &     5.8    &  7.97   \\
279    &   186    &    0.53    &     4.6    &  0.04    \\
280    &    76    &    0.05    &     4.0    &  0.24   \\
281    &   124    &    0.45    &     4.4    &  1.01   \\
283    &    16    &    1.67    &     4.6    &  5.57   \\
284    &   260    &    0.41    &     4.6    &  0.09   \\
285    &   132    &    0.14    &     4.2    &  1.16   \\
286    &   122    &    1.03    &     4.8    &  1.78   \\
287    &    16    &    0.16    &     3.6    &  1.98   \\
289    &   482    &    0.73    &     5.0    &  0.30   \\
290    &   850    &    0.39    &     5.0    &  4.69   \\
292    &    14    &    0.01    &     3.4    &  0.09    \\
293    &   850    &    0.89    &     5.4    &  4.89   \\
298    &   850    &    1.33    &     5.6    &  4.40   \\
299    &   132    &    1.40    &     5.0    &  4.13   \\
300    &   998    &    0.99    &     5.4    &  0.61   \\
301    &   850    &    0.79    &     5.2    &  2.31   \\
302    &   122    &    1.04    &     4.8    &  2.16   \\
304    &   850    &    1.05    &     5.4    &  7.25   \\
305    &   850    &    0.62    &     5.2    &  2.63   \\
306    &    16    &    0.67    &     4.0    &  0.16    \\
307    &   132    &    0.83    &     4.6    &  3.88   \\
308    &    16    &    0.81    &     4.0    &  1.54   \\
309    &   850    &    2.23    &     6.2    & 19.19   \\
310    &   124    &    0.50    &     4.4    &  0.47   \\
311    &   136    &    0.85    &     4.6    &  1.84   \\
313    &   122    &    0.84    &     4.6    &  5.09   \\
316    &     0    &    0.26    &     3.0    &  3.56   \\
317    &    16    &    1.08    &     4.2    &  4.14   \\
318    &    16    &    0.31    &     3.6    &  1.89   \\
324    &   122    &    0.18    &     4.2    &  3.84   \\
325    &    16    &    0.57    &     3.8    &  0.26   \\
326    &   126    &    0.27    &     4.2    &  0.03   \\
328    &   108    &    0.29    &     4.2    &  0.14   \\
331    &   124    &    0.00    &     4.0    &  0.03  \\
333    &    22    &    0.02    &     3.6    &  1.92   \\
335    &    24    &    0.45    &     3.8    &  1.65   \\
336    &   124    &    0.89    &     4.6    &  7.84   \\
340    &    16    &    0.13    &     3.6    &  0.25   \\
341    &   850    &    2.73    &     6.4    & 10.58   \\
342    &    16    &    1.23    &     4.2    &  3.81   \\
344    &   184    &    0.41    &     4.4    &  0.27   \\
348    &    50    &    0.99    &     4.4    &  0.07    \\
351    &   116    &    0.30    &     4.2    &  0.13  \\
352    &    16    &    0.59    &     3.8    &  0.25   \\
353    &   850    &    0.80    &     5.2    &  9.46   \\
354    &   850    &    0.70    &     5.0    &  0.66   \\
355    &   122    &    0.68    &     4.4    &  0.61   \\
359    &    16    &    0.83    &     4.0    &  3.52   \\
361    &    46    &    0.81    &     4.2    &  0.10    \\
362    &   122    &    0.90    &     4.6    &  4.21   \\
364    &    38    &    0.59    &     4.0    &  0.19   \\
365    &    16    &    0.23    &     3.6    &  4.39   \\
366    &   126    &    0.44    &     4.2    &  0.24   \\
369    &   122    &    0.73    &     4.4    &  1.14   \\
370    &   850    &    0.97    &     5.2    &  0.76   \\
371    &   124    &    2.58    &     5.6    & 14.58   \\
374    &   850    &    1.61    &     5.6    &  4.44   \\
377    &   850    &    1.59    &     5.6    &  6.07   \\
378    &   850    &    2.70    &     6.4    & 23.93   \\
379    &    16    &    2.94    &     5.2    & 10.26   \\
381    &    16    &    0.48    &     3.6    &  2.06   \\
383    &    16    &    1.33    &     4.2    &  5.22   \\
384    &   576    &    1.00    &     5.0    &  0.10  \\
386    &    16    &    0.84    &     3.8    &  1.50   \\
387    &    16    &    1.45    &     4.2    &  3.68   \\
388    &   212    &    0.00    &     4.0    &  0.12   \\
390    &   122    &    0.48    &     4.2    &  0.50   \\
392    &    16    &    0.34    &     3.6    &  0.64   \\
396    &    26    &    0.00    &     3.4    &  0.15   \\
397    &   122    &    0.58    &     4.2    &  0.84   \\
401    &    16    &    0.11    &     3.4    &  1.70   \\
406    &    16    &    0.27    &     3.4    &  4.89   \\
408    &    16    &    2.91    &     5.0    &  2.71   \\
409    &   124    &    1.91    &     5.0    &  1.98   \\
410    &   122    &    0.91    &     4.4    &  1.35   \\
414    &   850    &    1.81    &     5.6    &  2.43   \\
418    &    16    &    0.79    &     3.6    &  1.05   \\
419    &    16    &    2.40    &     4.6    & 16.45   \\
421    &   726    &    1.10    &     5.0    &  0.02   \\
423    &   850    &    1.58    &     5.4    & 18.21   \\
425    &    16    &    2.80    &     5.0    & 27.13   \\
427    &   122    &    1.59    &     4.8    &  1.49   \\
428    &    16    &    0.58    &     3.4    &  0.88   \\
429    &   122    &    0.80    &     4.2    &  0.11   \\
431    &   122    &    2.92    &     5.6    &  7.55   \\
438    &   850    &    2.99    &     6.4    &  6.13   \\
453    &    16    &    1.17    &     3.8    &  0.68   \\
461    &    16    &    2.82    &     4.6    & 13.81   \\
\end{longtable}
\textbf{Notes:} Results of the $\chi^2$ fitting of age (column 2), color excess (column 3), and mass (column 4) of the observed star clusters of NGC\,5248. The star cluster ID in column 1 is equal to that in the \citet{Maoz} electronic table.\\
\end{appendix}

\end{document}